\documentclass[nospthms,smallextended,final]{svjour3}
\usepackage{amssymb,amsmath,mathptmx}

\usepackage[utf8]{inputenc}
\usepackage[english]{babel}
\usepackage[T1]{fontenc}

\usepackage{amssymb,amsmath}
\usepackage{graphicx}
\usepackage[export]{adjustbox}
\usepackage{bm}
\usepackage{color}
\usepackage{subcaption}
\usepackage{hyperref}

\newcommand{\R}{\mathbb R}

\newcommand{\vc}[1]{\bm{#1}}
\newcommand{\vt}[1]{\mathsf{#1}}

\newcommand{\crs}{\mathcal{A}}

\newcommand{\rop}{{\mathsf{L}}}
\newcommand{\uop}{{\mathsf{U}}}
\newcommand{\rod}{{\mathcal{R}}}
\newcommand{\rg}{{\mathsf{SE}(3)}}
\newcommand{\ral}{{\mathfrak{se}(3)}}

\newcommand{\E}{{\mathcal{E}}}

\newcommand{\red}[1]{\textcolor{black}{#1}}

\begin{document}

\title{Importance and effectiveness of representing the shapes of Cosserat rods and framed curves as paths in the special Euclidean algebra}
\titlerunning{Shapes of Cosserat rods and framed curves as paths in the special Euclidean algebra}

\author{Giulio G. Giusteri \and Eliot Fried}
\institute{Mathematics, Mechanics, and Materials Unit\\ Okinawa Institute of Science and Technology Graduate University\\ 1919-1 Tancha, Onna, Okinawa, 904-0495, Japan\\
\email{giulio.giusteri@oist.jp; eliot.fried@oist.jp}}

\maketitle

\begin{abstract}
We discuss how the shape of a special Cosserat rod can be represented as a path in the special Euclidean algebra. 
By \emph{shape} we mean all those geometric features that are invariant under isometries of the three-dimensional ambient space.
The representation of the shape as a path in the special Euclidean algebra is intrinsic to the description of the mechanical properties of a rod, since it is given directly in terms of the strain fields that stimulate the elastic response of special Cosserat rods.
Moreover, such a representation leads naturally to discretization schemes that avoid the need for the expensive reconstruction of the strains from the discretized placement and for interpolation procedures which introduce some arbitrariness in popular numerical schemes.
Given the shape of a rod and the positioning of one of its cross sections, the full placement in the ambient space can be uniquely reconstructed and described by means of a base curve endowed with a material frame.
By viewing a geometric curve as a rod with degenerate point-like cross sections, we highlight the essential difference between rods and framed curves, and clarify why the family of relatively parallel adapted frames is not suitable for describing the mechanics of rods but is the appropriate tool for dealing with the geometry of curves.

\end{abstract}

\section{Motivation and main results}

Over the past century, rod theory has undergone a systematic development and has provided a platform for endless applications. We regard Antman's~\cite{Antman2005} presentation of the subject as the definitive reference regarding both the physical and mathematical foundations of the theory.
As for applications, a wealth of specialized references can be found. Here, we only mention models of elastic beams in structural engineering, studies of the shapes and instabilities of cables and chords, simulations of hair strands in computer graphics, and investigations of DNA supercoiling as evidence of the widespread usage of rod theory. In each of these applications what is used is the special Cosserat theory of rods, as introduced by the brothers Cosserat~\cite{Cos07,Cosserat1909} in 1907.
The first aim of the present paper is to put in evidence some features of the Lie algebraic structure that is implicit in the treatment of rod theory given by Antman~\cite{Antman2005}. That structure, while being only accessory to most analytical developments, is extremely relevant to the construction of discretization schemes able to capture some important traits of the theoretical framework.

We show that the shape of a rod, namely those features that are invariant under direct isometries of the three-dimensional ambient space, can be identified with a square-integrable path in the special Euclidean algebra. As explained in Section~\ref{sec:rod}, this emerges because the cross sections of a rod are assumed to be rigid and the special Euclidean group is the Lie group that describes the possible placements of a rigid body in three-dimensional space. By virtue of the tacit continuity assumptions of rod theory, a purely Lie algebraic description of the rod shape is available. 
The main feature of this approach is that information about the shape of a rod is not encoded in a description of what we see in the ambient space but instead stems from a description of the procedure that we must follow to redraw what we see. 

Such a representation of the rod shape, though not intuitive, appears to be extremely natural once it is recognized that it is defined in terms of the same strain fields that are most commonly used to describe the material response of the rod. 
We show that the six strain fields are the only degrees of freedom necessary to determine the shape of a rod (accompanied, of course, by a description of the cross sections as two-dimensional sets).

Significantly, the general variational approach devised by Schuricht~\cite{Sch02} to study the equilibria of nonlinearly elastic rods with topological constraints (and recently adopted by Giusteri, Lussardi \& Fried \cite{GiuLus17} to study the Kirchhoff--Plateau problem) is tacitly based on the same Lie algebraic representation of the rod shape.
Moreover, the role of the special Euclidean algebra is also essential in connection with the geometric mechanical concepts described, for instance, in the works by Simo, Marsden \& Krishnaprasad~\cite{SimMar88}, Simo, Posbergh \& Marsden~\cite{SimPos90}, Holm, Noakes \& Vankerschaver~\cite{HolNoa13}, and Eldering \& Vankerschaver~\cite{EldVan14} and with the $G$-strand equations discussed by Holm \& Ivanov \cite{HolIva14}. It should be noted, however, that these authors apply geometric concepts to study the dynamics of rods, whereas we focus our attention on the description of shapes.

Due to the basic role played by the strain fields, simulation strategies based on this representation offer an easier management of the relevant physical information. 
We present, in Section~\ref{sec:discretization}, a very intuitive and yet powerful discretization scheme, that generalizes to special Cosserat rods the approach devised by Bertails, Audoly, Cani, Querleux, Leroy \& L{\'e}v{\^e}que~\cite{BerAud06} for Kirchhoff rods. 
The major advantage of this approach is that, operating directly at the level of the Lie algebra, it is never necessary to interpolate between different elements of the special Euclidean group. 
Interpolation or discrete differentiation are usually necessary to retrieve differential information about the shape of a rod---information that is essential to compute the material response---from the placements in the ambient space of a finite number of cross sections of the rod.
Unfortunately, there is no unique way to reconstruct that information. 
By contrast, we introduce a finite-element discretization of the rod shape in which the essential information is always available and from which the placements in the ambient space of the cross sections are uniquely determined. 
In Section~\ref{sec:relax}, we illustrate the effectiveness of that discretization by solving boundary-value problems to find the equilibrium shapes of special Cosserat rods.

The same Lie algebraic construction applies to the theory of framed curves. The points of any such curve are endowed with a triad of orthonormal vectors that constitute a frame field varying along the curve. Framed curves have been used to study topological and geometric invariants and as basic models for describing the kinematics of slender bodies.
In this context, they are sometimes considered equivalent to special Cosserat rods, but this commingling should be avoided. Indeed, as the name suggests, the notion of framed curve rests upon the geometry of a curve as the basic constituent, according to which the frame field should be constructed. In contrast, the basic objects in rod theory are the material cross sections.

The second objective of this paper is thus to clarify the distinction between special Cosserat rods and framed curves.
By deriving, in Section~\ref{sec:framed-curve}, the theory of framed curves as a limiting case of the Cosserat theory, we show that the former theory is not adequate to describe the mechanics of rods, since it is incapable of tracking twisting and shearing deformations and completely neglects any effect due to the actual shapes of the cross sections. 
Even in those cases in which the frame along the curve is chosen to represent the material frame (and not merely determined by the curve geometry), the essential role accorded to the base curve makes it difficult to factor out global isometries. 
It also imposes viewing the strain fields as derived degrees of freedom, at odds with their primary role in the mechanical theory of rods.

Our derivation of the theory of framed curves highlights the relevance of the results presented by Bishop~\cite{Bis75} in 1975, which are still surprisingly ignored in some recent publications. We generalize his construction of relatively parallel adapted frames to the case of continuously differentiable regular curves. We show that the corresponding family of frame fields is uniquely determined by the geometric invariants of a generic curve. We also identify such geometric invariants with a square-integrable curvature field and a measure-valued torsion field, the regularity of which cannot be improved without imposing additional assumptions. We conclude by remarking that, when treating purely geometric questions surrounding space curves, relatively parallel adapted frames are the appropriate tool, and any use of the Frenet frame should be abandoned.

\section{Describing a thin rod}\label{sec:rod}

When modeling a filament or rod as a continuous body, we can mathematically express its slenderness by saying that, at any of its points, we can identify a direction in which the boundary of the body appears to be much farther away than it does in the two remaining orthogonal directions. If this is the case, we can represent the body as the collection of planar two-dimensional rigid bodies, named cross sections. The special Cosserat theory of rods (as presented, for instance, by Antman~\cite{Antman2005}) is predicated on the assumption that these cross sections are rigid and can only rotate or translate in space when the rod deforms.
It is then clear that the configuration of a special Cosserat rod (henceforth referred to simply as a rod) is fully described by assigning a family of two-dimensional sets, describing the {material cross sections}, and specifying how those sets are placed in three-dimensional (ambient) space. 
On the other hand, the shape of a rod is invariant under isometries of the ambient space and it is encoded in the relative placement of infinitesimally close cross sections.

For definiteness, we describe the family of cross sections, parametrized by $s$ in the interval $[0,L]$, as given by compact simply connected domains $\crs(s)$ of $\R^2$. It is important to clearly state a continuity assumption to make sure that any positioning of the collection of cross sections in space forms a continuous body. A first step toward guaranteeing continuity is to assume that the origin $\vc 0_2$ of $\R^2$ belongs to the interior of $\crs(s)$ for every $s$. Although the choice of $\vc 0_2$ is convenient, we emphasize that it is completely arbitrary. Using any other point of $\R^2$ is allowed and it is also possible to devise different conditions.

The placement in three-dimensional space of the cross section for each $s$ is fixed by assigning the image $\vc x(s)$ of the origin $\vc 0_2$ of the cross-sectional plane and the images $\vc d_1(s)$ and $\vc d_2(s)$ of a common orthonormal basis of $\R^2$ used to describe the cross sections. Since the cross sections are rigid, $\vc d_1(s)$ and $\vc d_2(s)$ together with $\vc d_3(s):=\vc d_1(s)\times\vc d_2(s)$ constitute an orthonormal basis for $\R^3$; that basis is referred to as the material frame at $s$. The second ingredient of the continuity assumption requires that the collection
\[
\big\{(\vc x(s),\vc d_3(s),\vc d_1(s),\vc d_2(s)):s\in[0,L]\big\}
\]
describe a continuous path in $(\R^3)^4\cong\R^{12}$.
Given this path,
the placement in three-dimensional space of the rod corresponds to the image of the set
\begin{equation*}
\Omega:=\big\{(s,\zeta_1,\zeta_2):s\in[0,L]\text{ and }(\zeta_1,\zeta_2)\in\crs(s)\big\}
\end{equation*}
through the map
\begin{equation}\label{eq:p}
\vc p(s,\zeta_1,\zeta_2):=\vc x(s)+\zeta_1\vc d_1(s)+\zeta_2\vc d_2(s).
\end{equation}

Assuming that path to be differentiable, we seek to identify an initial-value problem that describes how it is traced.
The initial conditions are obviously set by the given values of $\vc x(0)$, $\vc d_3(0)$, $\vc d_1(0)$, and $\vc d_2(0)$.
Taking into account that $\vc d_1(s)$, $\vc d_2(s)$, and $\vc d_3(s)$ are orthonormal, the relevant ordinary differential equations to be solved for $s\in(0,L)$ are
\begin{equation}\label{eq:Cauchy}
\left\{\begin{aligned}
&\vc x'(s)=v_3(s)\vc d_3(s)+v_1(s)\vc d_1(s)+v_2(s)\vc d_2(s),\\
&\vc d'_3(s)=u_2(s)\vc d_1(s)-u_1(s)\vc d_2(s),\\
&\vc d'_1(s)=-u_2(s)\vc d_3(s)+u_3(s)\vc d_2(s),\\
&\vc d'_2(s)=u_1(s)\vc d_3(s)-u_3(s)\vc d_1(s),
\end{aligned}\right.
\end{equation}
where a prime denotes differentiation with respect to $s$.

The strain fields $u_i$ and $v_i$, for $i=1,2,3$, have the following geometric interpretations. Indicating by $ds$ an infinitesimal increment of arclength, $u_i(s)$ represents the differential rotation about $\vc d_i(s)$ needed to bring the material frame at $s$ onto the material frame at $s+ds$; $u_1(s)$ and $u_2(s)$ thus concern flexural deformations of the collection of cross sections, while $u_3(s)$ is associated with twisting deformations. Meanwhile, $v_i(s)$ represents the differential translation in the direction of $\vc d_i(s)$ needed to bring the image of the origin at $s$ onto the image at $s+ds$; $v_1(s)$ and $v_2(s)$ thus concern shearing between adjacent cross sections, while $v_3(s)$ governs the differential distance between them, since $\vc d_3(s)$ is normal to the cross section at $s$. 

\begin{figure}
\centering

(a)
\includegraphics[width=0.44\textwidth,valign=t]{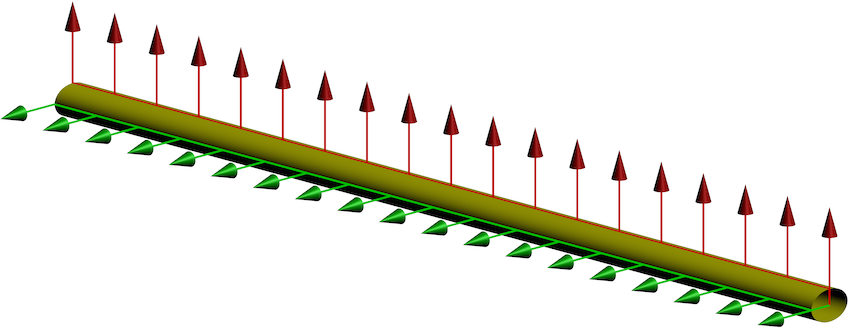}
\quad(b)
\includegraphics[width=0.44\textwidth,valign=t]{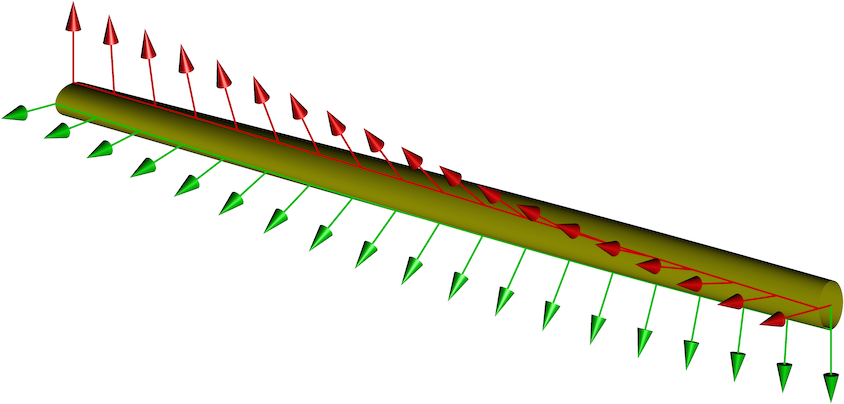}

\vspace{3mm}

(c)
\includegraphics[width=0.44\textwidth,valign=b]{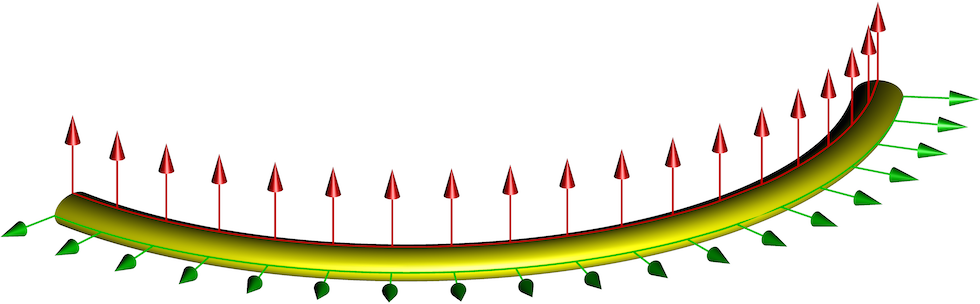}
\quad(d)
\includegraphics[width=0.44\textwidth,valign=b]{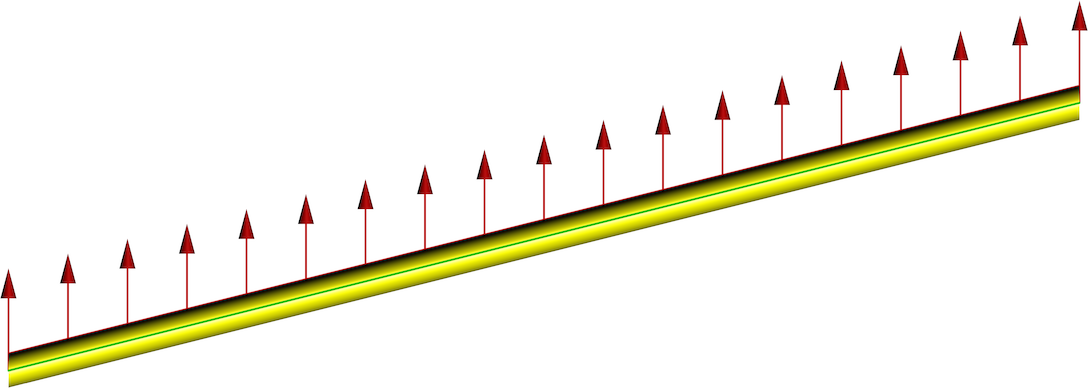}

\caption{The fields $u_i$ and $v_i$, $i=1,2,3$, define the shape of the rod. For clarity, we depict rods with uniform elliptical cross sections. The red and green arrows indicate the orientations of the material directors $\vc d_1$ and $\vc d_2$, respectively. The stretching field $v_3$ is identically equal to unity in all cases. (a) Straight rod: all strain fields except $v_3$ vanish identically. (b) Twisted rod: a constant non-vanishing value of $u_3$ produces a progressive rotation of the cross sections about $\vc d_3$. (c) Curved rod: a constant non-vanishing value of $u_1$ or $u_2$ produces curvature in the plane orthogonal to $\vc d_1$ or $\vc d_2$, respectively. (d) Sheared rod: a constant non-vanishing value for $v_1$ or $v_2$ produces shearing between adjacent cross sections.}\label{fig:rc}
\end{figure}

To better understand the effect of the various fields, it is useful to consider some particularly simple cases. First, to avoid the (physically undesirable) superposition of adjacent cross sections, it is necessary to require that $v_3$ satisfies the condition $v_3>0$. We can then take $v_3$ to be equal to unity and require that all other fields vanish and obtain a straight rod, in which the material frame simply translates in the fixed direction normal to the cross sections (Figure~\ref{fig:rc}a). On keeping $v_3$ equal to unity and assuming that at most one of the other fields is non-vanishing but uniform, we see that $u_3$ produces twisting by rotating the cross sections about $\vc d_3$ (Figure~\ref{fig:rc}b), while $u_1$ or $u_2$ produce curvature in the plane orthogonal to $\vc d_1$ or $\vc d_2$, respectively (Figure~\ref{fig:rc}c). With the alternative assumption that $v_1$ and $v_2$ take constant nonvanishing values, we obtain a shearing between adjacent cross sections, with a material frame that is simply translating in the fixed direction identified by $v_3\vc d_3+v_1\vc d_1+v_2\vc d_2$ (Figure~\ref{fig:rc}d).

In view of the differentiability assumption, the mapping $\vc x$ describes a differentiable curve in $\R^3$, parametrized by $s$ in the interval $[0,L]$. Such a base curve (called also midline, centerline, etc.) \red{gives a first approximation of the rod configuration} and it is most often taken as a starting point in the description of a rod. Nevertheless, we think that this point of view (albeit followed in our previous related publications) is misleading, since that curve is only expedient in describing the placement of the cross sections in space. We will return to analogies and differences between a rod and a framed curve in Section~\ref{sec:framed-curve}, but, to appreciate the immaterial nature of the base curve, it is enough to observe that it is possible to choose the sets describing the cross sections in such a way that the origin $\vc 0_2$ of $\R^2$ never belongs to $\crs(s)$, clearly showing that the points of the image of the base curve do not belong to the material points that constitute the rod as a continuous body. \red{A thorough analysis of the role of the base curve in rod theory is given by Antman \& Schuricht~\cite{AntSch03}, and analogous considerations for the case of shells were earlier provided by Naghdi~\cite{Nag75}.}

We now construct the vector field $\rod:[0,L]\to\R^{12}$ with the ordered components of $\vc x$, $\vc d_3$, $\vc d_1$, and $\vc d_2$ and introduce (denoting by $\vt O$ and $\vt I$ the $3\times 3$ matrices representing the null and identity endomorphisms of $\R^3$, respectively) the linear operator
\begin{equation}\label{eq:rop}
\rop(s):=
\begin{pmatrix}
\vt O & v_3(s)\vt I & v_1(s)\vt I & v_2(s)\vt I \\
\vt O & \vt O & u_2(s)\vt I & -u_1(s)\vt I \\
\vt O & -u_2(s)\vt I & \vt O & u_3(s)\vt I \\
\vt O & u_1(s)\vt I & -u_3(s)\vt I & \vt O 
\end{pmatrix}.
\end{equation}
In this way, the differential system \eqref{eq:Cauchy} can be rewritten as
\begin{equation}\label{eq:linsys}
\rod'=\rop\rod.
\end{equation}
Given the condition $\rod_0$ at $s=0$, and under mild measurability assumptions on the operator-valued map $\rop$, a unique solution of \eqref{eq:linsys} exists (see, for instance, the treatment by Hartman~\cite{Hartman1982}) and can be formally written as
\begin{equation}\label{eq:uop}
\rod(s)=\uop(s;0)\rod_0,
\end{equation}
where the operator $\uop(s_1;s_0)$ represents the propagator of the solution from the point $s_0$ to $s_1$.

From the construction above, we conclude that the shape of a rod, namely those features that are invariant under direct isometries of three-dimensional space, is fully encoded in the strain fields $u_i$ and $v_i$, $i=1,2,3$, that determine the operator $\rop$. At the same time, we see that the way in which a rod is rigidly translated and rotated in space depends solely on the initial conditions given by $\rod_0$.

\subsection{The Lie algebra and the Lie group associated with the rod description}\label{sec:as-paths}

Since a rod is defined by a collection of planar rigid cross sections continuously positioned in space, it is not surprising that the Lie algebra used to describe this system corresponds to the one needed to describe the positioning of rigid bodies in three spatial dimensions: it is the special Euclidean algebra $\ral$, which is associated with the special Euclidean group $\rg$ generated by rotations and translations of three-dimensional space.

Considering the definition of the operator $\rop$ in \eqref{eq:rop}, it is possible to identify useful representations for these structures.
It is immediately evident that there are six independent generators of $\ral$ that can be represented in $\mathsf{GL}_4(\R)$ by
\begin{gather*}
V_1=
\begin{pmatrix}
0&0&1&0\\
0&0&0&0\\
0&0&0&0\\
0&0&0&0\\
\end{pmatrix},\quad
V_2=
\begin{pmatrix}
0&0&0&1\\
0&0&0&0\\
0&0&0&0\\
0&0&0&0\\
\end{pmatrix},\quad
V_3=
\begin{pmatrix}
0&1&0&0\\
0&0&0&0\\
0&0&0&0\\
0&0&0&0\\
\end{pmatrix},
\\[4pt]
U_1=
\begin{pmatrix}
0&0&0&0\\
0&0&0&-1\\
0&0&0&0\\
0&1&0&0\\
\end{pmatrix},\quad
U_2=
\begin{pmatrix}
0&0&0&0\\
0&0&1&0\\
0&-1&0&0\\
0&0&0&0\\
\end{pmatrix},\quad
U_3=
\begin{pmatrix}
0&0&0&0\\
0&0&0&0\\
0&0&0&1\\
0&0&-1&0\\
\end{pmatrix}.
\end{gather*}
The structure constants
of $\ral$ are fixed by the commutation rules
\begin{equation}\label{eq:str-const}
[V_i,V_j]=0,\qquad[U_i,U_j]=-\varepsilon_{ijk}U_k,\qquad[V_i,U_j]=-\varepsilon_{ijk}V_k,
\end{equation}
for $i$, $j$, and $k$ ranging from $1$ to $3$, where $\varepsilon_{ijk}$ is the alternating Levi-Civita symbol.

Also the corresponding Lie group $\rg$ can be represented as a subgroup of $\mathsf{GL}_4(\R)$. Its elements can be obtained by applying the exponential map to the linear combinations of the generators of $\ral$.
Other representations of $\rg$ have been devised with the objective of reducing memory usage in computational settings (see, for instance, the presentation by Murray, Li \& Sastry~\cite{Murray1994}), but they are not needed in the present treatment.

The importance of the special Euclidean group $\rg$, as a subgroup of the affine group, for the discussion of motion and shape representations in computer graphics and geometric modeling is presented, for example, by Agoston~\cite{Agoston2005}.
The relevance of $\rg$ and the associated algebra $\ral$ to rod theory is acknowledged by Sanders~\cite{San10}, discussed in a review by Chirikjian~\cite{Chi10}, and exploited in beam modeling by Sonneville, Cardona \& Br\"uls~\cite{SonCarBru14,SonCar14}. 
These authors base their approaches on representing a rod through elements of the group $\rg$, but we propose that the algebra $\ral$ provides a representation that is more naturally and directly related to the shape of a rod.

We have already observed that a full description of the placement of a rod in space corresponds to a continuous path $\{\rod(s):s\in[0,L]\}$ in $\R^{12}$, accompanied by a description of the material cross sections, since these fully determine the placement map $\vc p$. Based on the decomposition \eqref{eq:uop} of $\rod(s)$ as the action of the propagator $\uop(s;0)$ on the initial point $\rod_0$, it is possible to factor out global rigid-body motions, encoded in $\rod_0$, and identify the corresponding equivalence class of placements with the path $\{\uop(s;0):s\in[0,L]\}$ in $\mathsf{GL}_{12}(\R)$. 
Specifically, since the operator $\uop(s;0)$ belongs, for any $s$, to a representation of $\rg$ within $\mathsf{GL}_{12}(\R)$, we can identify the placement of the rod modulo rigid transformations with the continuous path $\{\uop(s;0):s\in[0,L]\}$ in $\rg$. It is also immediately evident that, having obtained $\uop(s;0)$ by solving \eqref{eq:linsys}, the essential information encoding the shape of the rod can be identified with the possibly discontinuous path $\{\rop(s):s\in[0,L]\}$ in $\ral$.

A rod is sometimes described as the juxtaposition of a path in $\R^3$ (the base curve) and a path in $\mathsf{SO}(3)$ representing the collection of material frames. This point  of view\red{, popularized by the works of Simo, Marsden \& Krishnaprasad~\cite{SimMar88} and Simo, Posbergh \& Marsden~\cite{SimPos90},} does not seem to advance the objective of distinguishing between the shape of the rod and its placement in space; indeed it is akin to choosing the first three components of $\rod(s)$ and the rotational part of $\uop(s;0)$ to describe the rod and thereby introducing an unnecessary asymmetry.

All of the mentioned identifications---which exploit either $\rod$, $\uop(\cdot;0)$, or $\rop$---are relevant to the construction of computational schemes for the simulation of rods and different discretized representations of a rod can be interpreted as different ways to discretize those paths. {Effective discretizations of rods to model slender bodies have been developed, among others, by Cao, Liu \& Wang~\cite{CaoLiu06}, Spillmann \& Teschner~\cite{SpiTes07}, Bergou, Wardetzky, Robinson, Audoly \& Grinspun~\cite{BerWar08}, Bergou, Audoly, Vouga, Wardetzky \& Grinspun~\cite{BerAud10}, Audoly, Clauvelin, Brun, Bergou, Grinspun \& Wardetzky~\cite{AudCla13}, Jung, Leyendecker, Linn and Ortiz~\cite{JunLey11}, Lang, Linn \& Arnold~\cite{LanLin11}, and Linn~\cite{Lin16}. A vast literature also exists in which rod theory is applied to the computational mechanics of beams. These approaches are characterized by the fact that translational and rotational degrees of freedom are often considered separately and the beam shape is reconstructed by means of interpolation procedures. A selection of methods can be found in the works by Simo \& Vu-Quoc~\cite{SimVuQ86}, Borri \& Bottasso~\cite{BorBot94}, Ibrahimbegovi\'c~\cite{Ibr95}, Betsch \& Steinmann~\cite{BetSte02}, Meier, Popp \& Wall~\cite{MeiPop14,MeiPop15}, Ga\'ce\v{s}a \& Jeleni\'c~\cite{GacJel15}, Bauer, Breitenberger, Philipp, W\"uchner \& Bletzinger~\cite{BauBre16}, Yilmaz \& Omurtag~\cite{YilOmu16}, and Zupan \& Zupan~\cite{ZupZup16}.}

In all the foregoing examples, the discretization is performed at the level of either $\rod$ or $\uop(\cdot;0)$, that is, by considering the placement of the rod in space.
An important exception to this general trend can be found in the works by Zupan \& Saje~\cite{ZupSaj03,ZupSaj06}, \v{C}e\v{s}arek, Saje \& Zupan~\cite{CesSaj13} (mainly addressing linearized beam equations),  Su \& Cesnik~\cite{SuCes11}, and Schr\"oppel \& Wackerfu{\ss}~\cite{SchWac16}. There, discretization is performed at the level of Lie algebraic fields, called strains, but nodal values are of the essence and interpolation schemes are again needed to reconstruct the shape of a rod.

In Section~\ref{sec:discretization}, we introduce a discretization of the shape of a rod viewed as a path in the algebra $\ral$, as defined by $\rop$. This generalizes to special Cosserat rods the approach used by Bertails, Audoly, Cani, Querleux, Leroy \& L{\'e}v{\^e}que~\cite{BerAud06} for Kirchhoff rods and can be viewed as a bridge between the methods of Sanders~\cite{San10}, Chirikjian~\cite{Chi10}, and Sonneville, Cardona \& Br\"uls~\cite{SonCarBru14,SonCar14}, based on the special Euclidean group, and the aforementioned ones, based on Lie algebraic quantities. A distinguishing feature of the present approach is that it obviates the need for any interpolation associated with the reconstruction of the shape of a rod from a finite sampling of its placement in space.

\subsection{Constraints on the placement and on the shape of a rod}\label{sec:constraints}

We consider two classes of constraints: constraints concerning how a rod is positioned in space and constraints relative to the shape of a rod, usually termed \emph{internal} constraints.
Internal constraints are more easily represented as conditions on the path traced by the operator $\rop$ in $\ral$, whereas constraints on the placement of the rod are nicely enforced on the path given by $\rod$ in $\R^{12}$.

\subsubsection{Placement constraints}

The most prominent examples of placement constraints are the clamping conditions that indicate how the ends of a rod are held in space. These take the linear form
\begin{equation}\label{eq:clamping}
\rod(0)=\rod_0\qquad\text{and}\qquad \rod(L)=\rod_L,
\end{equation}
where $\rod_0$ and $\rod_L$ are given vectors in $\R^{12}$. Fixing $\rod(0)$ amounts to imposing both the positions of the ends of a rod and the orientations of the extremal cross sections. In particular, the tangent vector to the base curve is also fixed, showing that \eqref{eq:clamping} corresponds to what is usually termed clamping.

It is possible to express the clamping conditions in terms of $\rod_0$ and of the path traced by $\rop$ in $\ral$ only by means of the nonlinear and nonlocal expression of $\uop$ in terms of $\rop$. This shows that enforcing the clamping conditions can be a delicate issue when this representation of a rod is used. Moreover, the relation between $\uop$ and $\rop$ can be made explicit only in particular cases. It is fortunate that those cases can be exploited to set up computational schemes, as we will show in Section~\ref{sec:discretization} below.

Notably, the foregoing clamping conditions can also be used to describe closed rods and they can be adapted, as discussed at the end of Section~\ref{sec:framed-curve}, to express the closure constraint when dealing with framed curves.

\subsubsection{Internal constraints}

Regarding internal constraints, of great importance are those leading to the classical Kirchhoff~\cite{Kir59} model. (See also the interesting account by Dill~\cite{Dil92}.) This model
adds two assumptions to those of the special Cosserat theory: (i) absence of shearing between adjacent cross sections, which amounts to setting, for every $s\in[0,L]$,
\begin{equation}\label{eq:unshearability}
v_1(s)=0\qquad\text{and}\qquad v_2(s)=0
\end{equation}
in \eqref{eq:Cauchy}$_1$, and (ii) inextensibility of the base curve, which can be achieved by setting, for every $s\in[0,L]$,
\begin{equation}\label{eq:inextensibility}
v_3(s)=1.
\end{equation}
Note that, in the representation of the rod shape within $\ral$, the unshearability and inextensibility constraints \eqref{eq:unshearability}--\eqref{eq:inextensibility} are both linear.

The primary consequences of Kirchhoff's assumptions are that, since \eqref{eq:Cauchy}$_1$ now takes the form $\vc x'=\vc d_3$, the third director of the material frame at $s$ corresponds to the tangent vector to the base curve and $s$ is the arc-length parameter of that curve.
With this, we can relate the fields $u_1$ and $u_2$ to two {flexural densities}, say $\kappa_1$ and $\kappa_2$, which are the components of the curvature vector $\vc t'$ in the directions of $\vc d_1$ and $\vc d_2$, respectively, and we can identify $u_3$ with the twisting, say $\omega$. Substituting the identifications
\[
\vc d_3=\vc t,\qquad u_1=-\kappa_2,\qquad u_2=\kappa_1 ,\qquad\text{and}\qquad u_3=\omega
\]
in \eqref{eq:Cauchy}, we find that the differential equation describing the placement of a Kirchhoff rod takes the form
\begin{equation}\label{eq:Cauchy-K}
\left\{\begin{aligned}
&\vc x'(s)=\vc t(s),\\
&\vc t'(s)=\kappa_1(s)\vc d_1(s)+\kappa_2(s)\vc d_2(s),\\
&\vc d'_1(s)=-\kappa_1(s)\vc t(s)+\omega(s)\vc d_2(s),\\
&\vc d'_2(s)=-\kappa_2(s)\vc t(s)-\omega(s)\vc d_1(s),
\end{aligned}\right.
\end{equation}
for $s$ in $(0,L)$.

Evidently, there is a considerable simplification in the model, since only three scalar fields determine the shape of a rod constrained in accord with \eqref{eq:unshearability} and \eqref{eq:inextensibility}. Perhaps surprisingly, however, there is absolutely no simplification in the Lie algebra and group necessary to describe the system. Indeed, due to the commutation relation $[V_i,U_j]=-\varepsilon_{ijk}V_k$, the unavoidable presence of the generator $V_3$ in the algebra associated with \eqref{eq:Cauchy-K} requires that $V_1$ and $V_2$ both remain in the picture. Hence, $\ral$ and $\rg$ are again the relevant mathematical structures to be considered.

\section{Discretizing the rod shape in $\ral$}\label{sec:discretization}

In this section, we introduce a discretization of the shape of a rod based on its representation as a path in the special Euclidean algebra $\ral$. We also discuss the advantages and limitations of this approach, with particular reference to variational descriptions of the rod elasticity.
For the special case of a Kirchhoff rod, this discretization scheme reduces to the one used by Bertails, Audoly, Cani, Querleux, Leroy \& L{\'e}v{\^e}que~\cite{BerAud06}. We moreover discuss the connection between {our} approach and the interpolation of affine transformations introduced very recently by Kaji \& Ochiai~\cite{KajOch16} in the context of computer graphics applications.

The most important feature of our perspective is that it does not rest on discretizing the placement of a rod in space. We instead discretize the shape of the rod. The placement in space is uniquely determined by the shape of a rod and the placement of one of its ends and it can be easily reconstructed. The converse is not true, and this shows the major advantage of the present method. Indeed, there is no unique way to reconstruct the shape of a rod from a discretization of its placement in space, as testified by the large number of interpolation techniques proposed in the literature (reviewed, for instance, by Romero~\cite{Rom08} and Bauchau \& Han~\cite{BauHan14}).

The starting point for the scheme is the observation that the solution of a first-order linear ordinary differential equation with constant coefficients can be represented explicitly using the matrix exponential map. As we already mentioned, the representation is not explicit in the general case of non-constant coefficients, but it remains explicit for piecewise constant coefficients.

We can then introduce a partition $P^{(N)}=\{0=s_0,s_1,\ldots,s_N=L\}$ of the interval $[0,L]$ and approximate the scalar fields $u_i$ and $v_i$, $i=1,2,3$, as piecewise constant (and right-continuous) on the intervals defined by $P^{(N)}$. Those fields fully describe the shape of a rod and their approximation corresponds to the definition of of an operator field $\rop$ that takes the constant value $\rop(s_{k-1})$ on the whole interval $[s_{k-1},s_k)$, for $k=1,\ldots,N$.
Consequently, on each interval 
we have
\begin{equation}\label{eq:propk}
\exp\bigg(\int_{s_{k-1}}^{s_k}\rop(t)\,dt\bigg)=\exp\bigg(\delta s_{k-1}\rop(s_{k-1})\bigg)=:\uop_k,
\end{equation}
having set $\delta s_{k-1}:=s_k-s_{k-1}$.

Relation \eqref{eq:propk} uniquely defines discrete propagators $\uop_k$, $k=1,\ldots,N$, that can be used to reconstruct the (discretized) placement of the rod cross sections in space, by giving an initial cross section as a vector $\rod_0$ in $\R^{12}$ and successively applying equation~\eqref{eq:uop}. We then see that a piecewise constant finite-element approximation of the operator field $\rop$---which is a path in $\ral$---provides a uniquely defined approximation of the placement of a rod through a discretization of its shape.

It is now worth commenting on the connection between our approach and the work of Kaji \& Ochiai~\cite{KajOch16}. As a particular case, their results provide a parametrization of the group $\rg$ in terms of the algebra $\ral$. That parametrization can be used to describe the cross sections that represent the nodes of a discretization of the placement of a rod. This is only part of the information contained in the shape of a rod, since no strategy for going from one cross section to another is specified. The function ``Blend'' is used by Kaji \& Ochiai~\cite[Sect.~5.2]{KajOch16} to interpolate between two cross sections in a way which is consistent with additional information about the ``shape'' of the interpolation. For instance, they are free to prescribe the total twist accumulated between two cross sections. Although their tool is clearly very flexible and useful for graphics manipulations, their approach cannot be used to render a rod without providing additional information about its shape. Our perspective differs because we take the discretized shape of a rod as primitive information and then uniquely reconstruct the rod placement in space. Whereas Kaji \& Ochiai~\cite{KajOch16} use the elements of $\ral$ to parametrize $\rg$, we use piecewise constant paths in $\ral$ to encode the shape of a rod and reconstruct elements of $\rg$, such as $\uop_k$, only when necessary. 

It should now be clear that the present discretization scheme is particularly useful when the information about the shape of a rod (and not its placement in space) is of central importance. This is particularly true whenever elastic beams or filaments are modeled by variational methods. Such methods always require the definition of an energy functional, the form of which characterizes the elastic response of the rod, and the main contribution to the stored elastic energy of a rod is always related to its shape. In this context, the need to reconstruct the information about the shape from the discretized placement is a potential source of difficulty.
On the contrary, our construction is directly expressed in terms of shape parameters, the strain fields $u_i$ and $v_i$, $i=1,2,3$, that uniquely determine the placement. In simple words, we always know how we go from a cross section to the adjacent one and this determines the elastic energy.

Another advantage of the present scheme is that, in each of the discretization intervals, the portions of a rod are generic helical segments. Hence, we can describe without any approximation certain curved configurations, as long as their shapes correspond to piecewise constant paths in $\ral$. Moreover, the internal constraints of unshearability~\eqref{eq:unshearability} and inextensibility~\eqref{eq:inextensibility} discussed in Section~\ref{sec:constraints} can be imposed exactly because they are compatible with piecewise constant values of the fields $v_1$, $v_2$, and $v_3$. 

On the contrary, a major difficulty in our approach arises from the clamping conditions~\eqref{eq:clamping}$_2$ at $s=L$, which constitute a highly nonlinear constraint. This is clear from the expression of~\eqref{eq:clamping}$_2$ in terms of the discrete propagators $\uop_k$, which reads
\begin{equation}\label{eq:exp-con}
\prod_{k=1}^N\uop_k=\vt B,
\end{equation}
where $\vt B$ is a linear transformation that maps the initial condition $\rod_0$ to the final condition $\rod_L$ and successive operators are multiplied from the left.

\subsection{Examples}

Here, we present a few examples of discretized rod shapes and renderings of the corresponding placements in three-dimensional space. To avoid situations in which the cross sections are trivially superimposed (and which thus are of no physical interest) we always assume that $v_3(s)=1$ for each value of $s$. Although this is not enough to guarantee non-interpenetration of matter, it rules out some trivial cases where this occurs.
For the clamping condition at $s=0$, we always assume that the base curve starts at the origin and that the material frame is aligned with a fixed orthonormal reference frame. We will first present ``exact approximations'', namely cases in which the strain fields are uniform on the entire interval $[0,L]$.

\begin{figure}
\centering

(a)
\includegraphics[width=0.42\textwidth,valign=b]{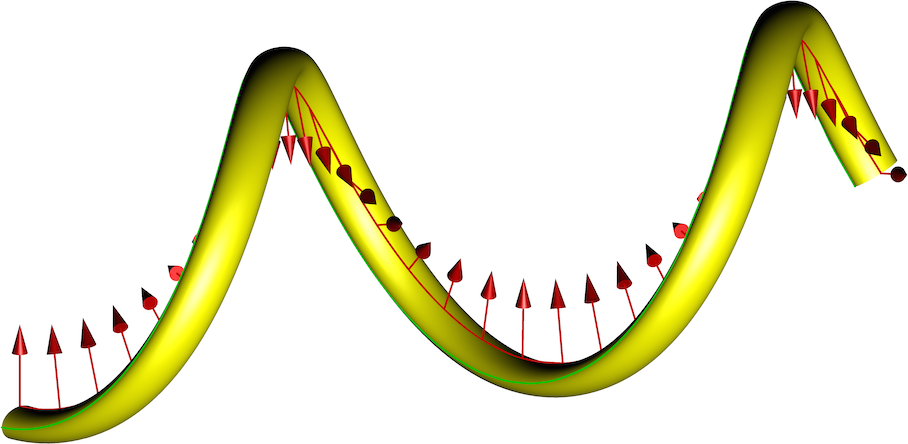}
\hspace{0.02\textwidth}
\quad(b)
\includegraphics[width=0.44\textwidth,valign=b]{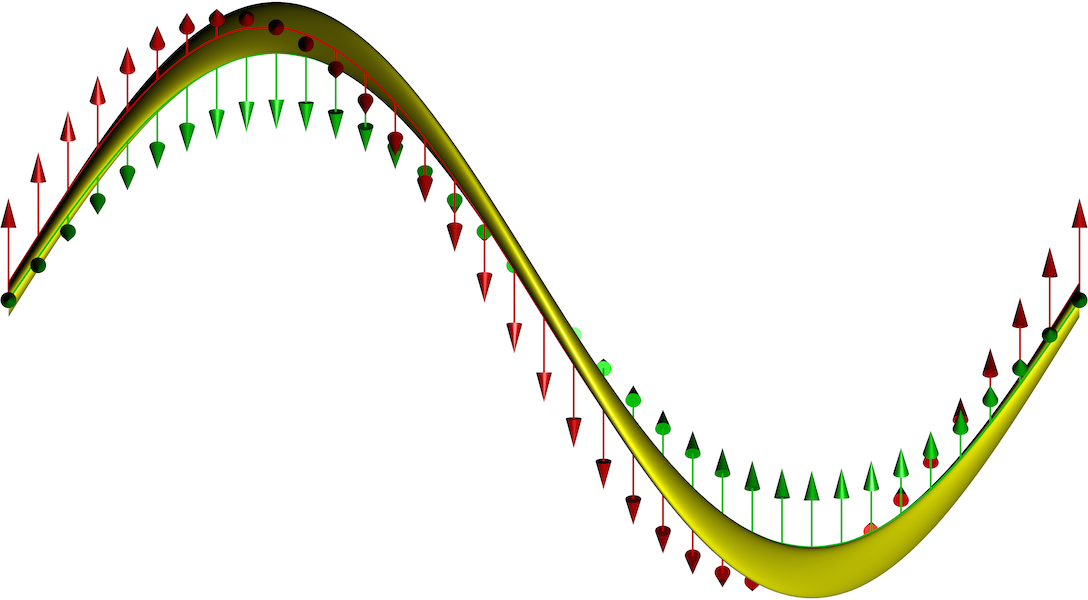}

\vspace{3mm}

(c)
\hspace{0.02\textwidth}
\includegraphics[width=0.38\textwidth,valign=b]{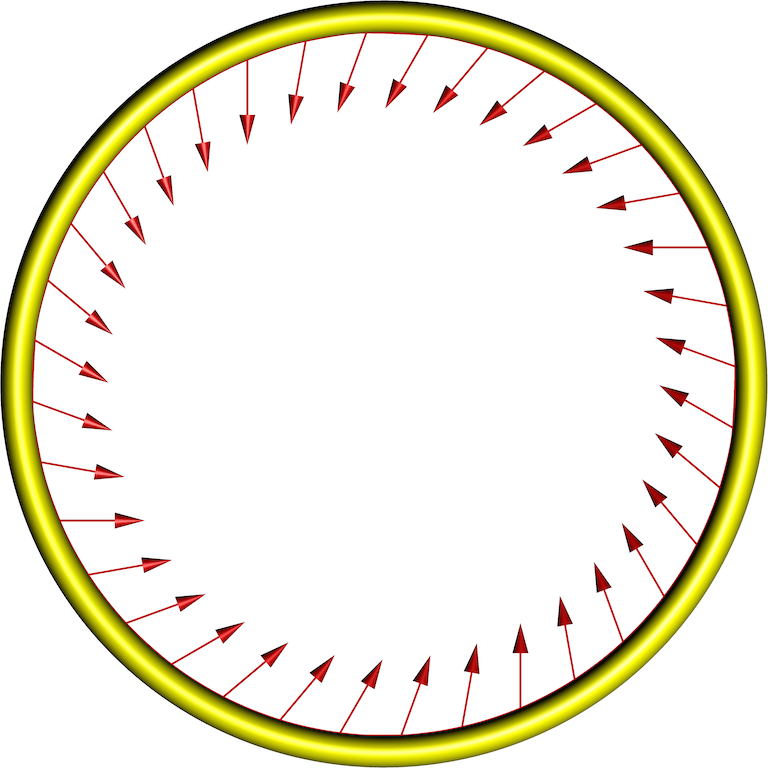}
\hspace{0.04\textwidth}
\quad(d)
\hspace{0.04\textwidth}
\includegraphics[width=0.28\textwidth,valign=b]{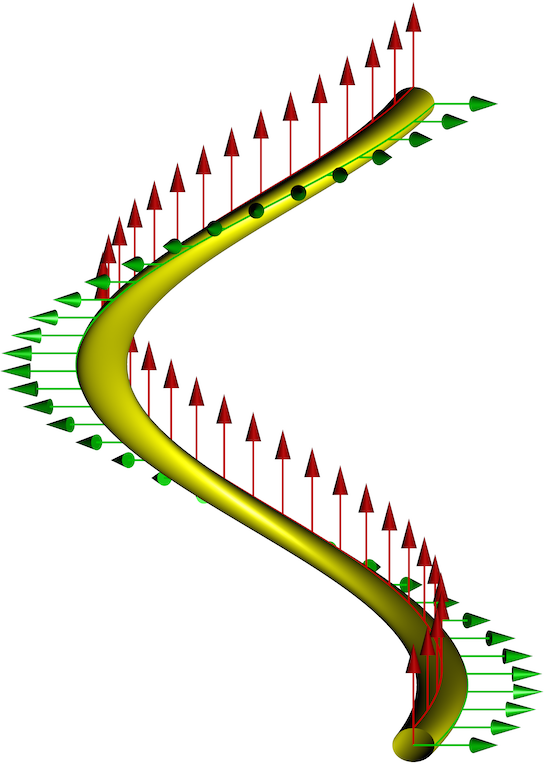}
\hspace{0.12\textwidth}

\caption{A single element in the shape discretization can represent a helical segment with possibly sheared cross sections. Without sharing, we need non-vanishing constant values of the twisting $u_3$ and the flexural densities $u_1$ or $u_2$ to obtain a helical segment (a).
Combining twisting and shearing, we obtain a helical segment (b) without flexural deformations. Avoiding twisting, we can combine shearing with flexural modes in the same direction, obtaining a sheared loop (c), or in the orthogonal direction, obtaining another helix (d).}\label{fig:helix-shear}\label{fig:helix0}
\end{figure}

First, we consider the special case of a Kirchhoff rod, for which unshearability and inextensibility are assumed, namely for which $v_1$ and $v_2$ vanish identically and $v_3$ is identically equal to unity:
\begin{itemize}
\item[$\bullet$] Taking $u_1(s)=c_1\neq 0$, $u_2(s)=c_2\neq 0$, and $u_3(s)=0$ for each $s$ in $[0,L]$, we arrive at a description of a twist-free circular arc with scalar curvature $\kappa$ of the base curve given by $\kappa=(c_1^2+c_2^2)^{1/2}$ (Figure~\ref{fig:rc}c).
\item[$\bullet$] Taking $u_1(s)=u_2(s)=0$ and $u_3(s)=c_3\neq 0$ for each $s$ in $[0,L]$, we arrive at a description of a straight rod with total twist $T$ given by $T=c_3L$ (Figure~\ref{fig:rc}b).
\item[$\bullet$] Taking $u_1(s)=c_1\neq 0$, $u_2(s)=c_2\neq 0$, and $u_3(s)=c_3\neq 0$ for each $s$ in $[0,L]$, we arrive at a description of a helical segment (Figure~\ref{fig:helix0}a).
\end{itemize}
We also consider examples in which shear is included:
\begin{itemize}
\item[$\bullet$] Taking constant values of $v_1$, $v_2$, and $v_3$ and vanishing $u_1$, $u_2$, and $u_3$, we arrive at a description of a straight sheared rod (Figure~\ref{fig:rc}d).
\item[$\bullet$] Taking non-vanishing constant values for $v_1$, $v_3$, and $u_3$, we arrive at a description of a helix without flexural deformations (Figure~\ref{fig:helix-shear}b).
\item[$\bullet$] Taking non-vanishing constant values for $v_1$, $v_3$, and $u_2$, we arrive at a description of a sheared circular arc (Figure~\ref{fig:helix-shear}c).
\item[$\bullet$] Taking non-vanishing constant values for $v_1$, $v_3$, and $u_1$, we arrive at a description of another helical shape (Figure~\ref{fig:helix-shear}d).
\end{itemize}

Any rod represented with our discretization is an assembly of segments with shapes of the kind described. The strain fields need not be continuous at the joints, so it is possible to exactly describe a rod formed, for instance, by two circular arcs lying in distinct planes using a partition of $[0,L]$ in just two subintervals.
It should be noted, however, that whereas the tangent field to the base curve of a Kirchhoff rod is continuous by construction---the base curve is a regular curve---that field may be discontinuous for a shearable Cosserat rod.

\begin{figure}
\centering

\begin{subfigure}[c]{.58\textwidth}
\includegraphics[width=\textwidth]{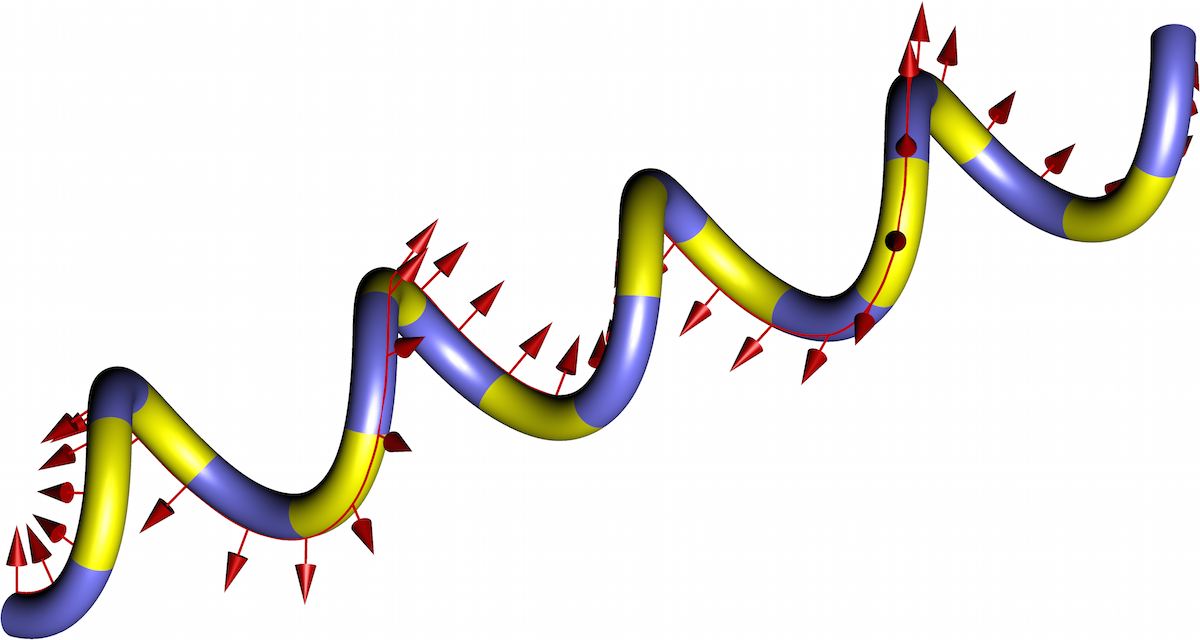}
\end{subfigure}
\begin{subfigure}[c]{.40\textwidth}
\hspace{0.08\textwidth}
\includegraphics[width=0.92\textwidth]{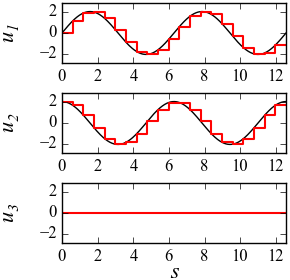}
\end{subfigure}

\vspace{2mm}

\begin{subfigure}[c]{.58\textwidth}
\includegraphics[width=\textwidth]{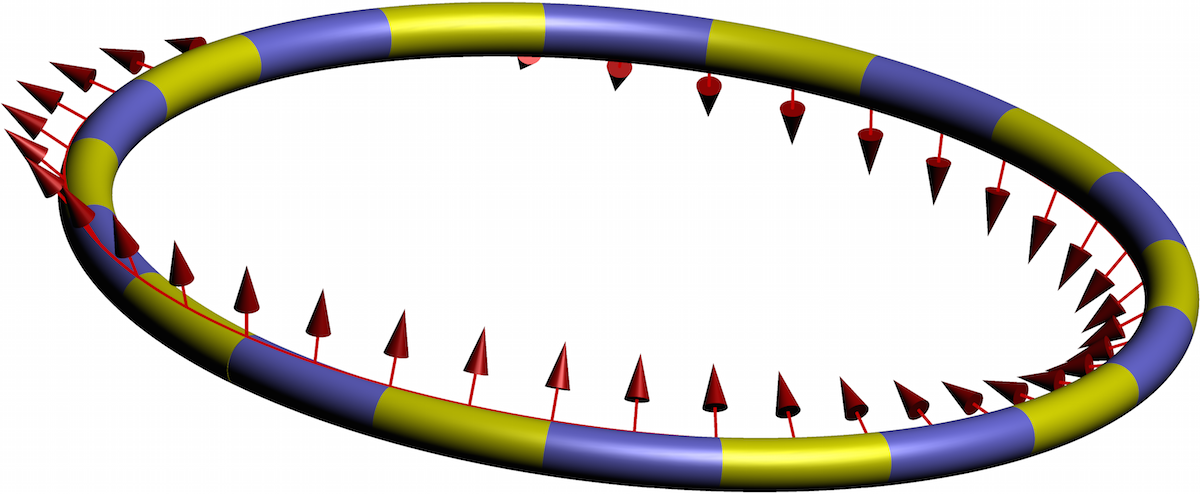}
\end{subfigure}
\begin{subfigure}[c]{.4\textwidth}
\hspace{0.08\textwidth}
\includegraphics[width=0.92\textwidth]{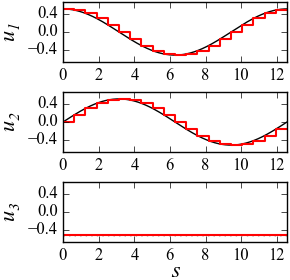}
\end{subfigure}

\caption{The piecewise constant approximation of strain fields makes it possible to describe any rod configuration. We see the rendering and the strain fields of a twist-free helical rod (top) and of a circular loop with constant twisting (bottom). Different rod elements are identified by the alternating coloring and the arrows indicate the orientation of the director $\vc d_1$. The twisting field, being constant, can be exactly captured by the discretization, while the flexural densities (being trigonometric functions, as needed to ensure a constant scalar curvature) must be approximated. Black lines show the theoretical values of the fields.}\label{fig:helix1}
\end{figure}

Two interesting shapes of Kirchhoff rods that cannot be exactly represented are a twist-free helical rod (Figure~\ref{fig:helix1}, top panels) and a circular loop with constant twisting (Figure~\ref{fig:helix1}, bottom panels). The rods presented in Figure~\ref{fig:helix1}, obtained using twenty-one rod elements, show how a generic rod configuration can be described within this setting, while highlighting the limits of a piecewise constant approximation of the strain fields (Figure~\ref{fig:helix1}, right panels). Indeed, twist-free elements are circular arcs, so a twist-free helical rod is represented as a collection of circular arcs. Meanwhile, a curved element with non-vanishing twisting has a helical (hence non-planar) base curve, so a circular loop with constant twisting is necessarily composed by helical segments and has a non-planar base curve. 
Clearly, the geometric error introduced by such approximations diminishes with refinement of the discretization and is inversely proportional to the number of rod elements.

\section{Application to shape relaxation}\label{sec:relax}

The representation and discretization of the rod shape presented in the previous sections is particularly effective in dealing with shape relaxation problems. 
Here we provide a few examples. Our objective is to illustrate the main advantage of the proposed approach, namely that operating directly at the level of the strain fields makes it possible to easily treat problems in which all the deformation modes of a rod are combined and also to represent with a small number of elements nontrivial curvilinear configurations.
Since our emphasis is on the shape representation and not on the solution procedure, we employ a simple and reliable gradient flow algorithm to find equilibrium configurations, but we do not aim to optimize the efficiency of the implementation.

We introduce a simple energy that penalizes deviations from the intrinsic shape of the rod, as encoded by the fields $\bar{u}_i$ and $\bar{v}_i$, $i=1,2,3$:
\begin{multline}\label{eq:energy}
\E(u_1,u_2,u_3,v_1,v_2,v_3):=\frac{1}{2}\int_0^{L}a_1(s)|u_1(s)-\bar{u}_1(s)|^2\,ds\\
+\frac{1}{2}\int_0^{L}\Big(a_2(s)|u_2(s)-\bar{u}_2(s)|^2+a_3(s)|u_3(s)-\bar{u}_3(s)|^2\Big)\,ds\\
+\frac{1}{2}\int_0^{L}\Big(b_1(s)|v_1(s)-\bar{v}_1(s)|^2+b_2(s)|v_2(s)-\bar{v}_2(s)|^2\Big)\,ds\\
+\frac{1}{2}\int_0^{L}b_3(s)\big(|v_3(s)-\bar{v}_3(s)|^2-2\epsilon\log v_3(s)\big)\,ds\,.
\end{multline}
The flexural rigidities $a_1>0$ and $a_2>0$, twisting rigidity $a_3>0$, shear rigidities $b_1>0$ and $b_2>0$, and the stretching rigidity $b_3>0$ are material fields that determine the strength of the elastic response to the corresponding deformations. The logarithmic perturbation in the final term (with $0<\epsilon\ll 1$ arbitrary) is introduced to ensure that $v_3$ remains always strictly positive, preventing total contraction.
Since no coupling between flexural and twisting modes is present, this is not the most general energy functional, even in the quadratic case. Nevertheless, our approach is easily applied to other choices of the energy functional.

The intrinsic shape defined by the fields $\bar{u}_i$ and $\bar{v}_i$, $i=1,2,3$, obviously represents the unique minimizer of the functional $\E$ in the absence of additional constraints on the strain fields (up to a correction to $\bar{v}_3$ of order $\epsilon$). However, if clamping conditions are imposed at both ends of the rod, those conditions constitute a nonlinear constraint that defines the submanifold of $\ral$ of admissible strain fields. If the intrinsic shape is not compatible with the clamping conditions, then the minimizer of $\E$ is no longer obvious.  We can also encounter situations in which multiple local minima are present, since the constraint is not convex.

To approximate energy minima numerically, we apply a gradient flow scheme to the functional $\E$ while exploiting the discretization of strain fields discussed in the previous section. Given that the constraint manifold is, in general, nonlinear, a strategy for enforcing the constraint at each iteration is needed. Since the exponential map from $\ral$ to $\rg$ involved in the definition of the clamping constraint admits a closed form expression (see, for instance, the work by Kaji \& Ochai~\cite{KajOch16}), it is possible to explicitly compute its gradient and apply a manifold projection method, as discussed by Hairer~\cite{Hairer2006}.

\begin{figure}
\centering

\includegraphics[width=\textwidth]{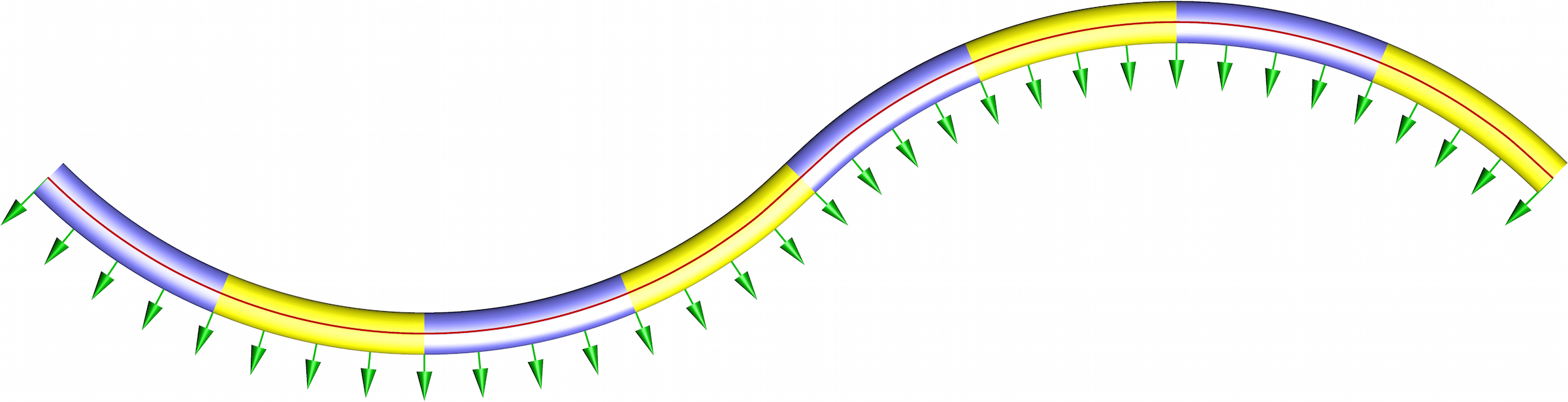}

\vspace{8mm}

\includegraphics[width=\textwidth]{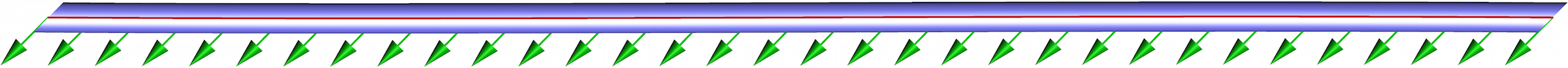}

\caption{The shape relaxation of a shearable and extensible rod with clamped ends can be reproduced. The rod is discretized using eight elements and the initial configuration (top) relaxes to the equilibrium configuration (bottom). The equilibrium configuration corresponds to a uniformly sheared rod and differs from the intrinsic shape due to the presence of clamping constraints at both ends.}\label{fig:rod-relax}
\end{figure}

In what follows, we assume, for simplicity, uniform (namely, $s$-independent) cross sections and rigidities, with a noncircular profile of the cross sections that translates into unequal values of the rigidities.
In our first example we have a very simple intrinsic shape, with vanishing curvature, twisting, and shearing, but with uniform stretching density $\bar{v}_3=1$. Regarding the rigidities, we emphasize the resistance to bending and stretching by setting $b_1=b_2=b$, $a_1/b=a_2/b=b_3/b=100$, and $a_3/b=10$. (For the logarithmic perturbation, we set $\epsilon=10^{-6}$.) The initial configuration is formed by two circular arcs lying in the plane orthogonal to the constant director $\vc d_1$ (Figure~\ref{fig:rod-relax}, top). In this configuration the rod is bent and stretched in comparison to its intrinsic shape. When relaxed, the configuration reaches an equilibrium in which the stretching density approaches the intrinsic value and curvature essentially disappears, in favor of the less costly shearing, which is needed to comply with the clamping conditions (Figure~\ref{fig:rod-relax}, bottom).

\begin{figure}
\centering

(a)\hspace{0.01\textwidth}
\includegraphics[scale=1.8,valign=t]{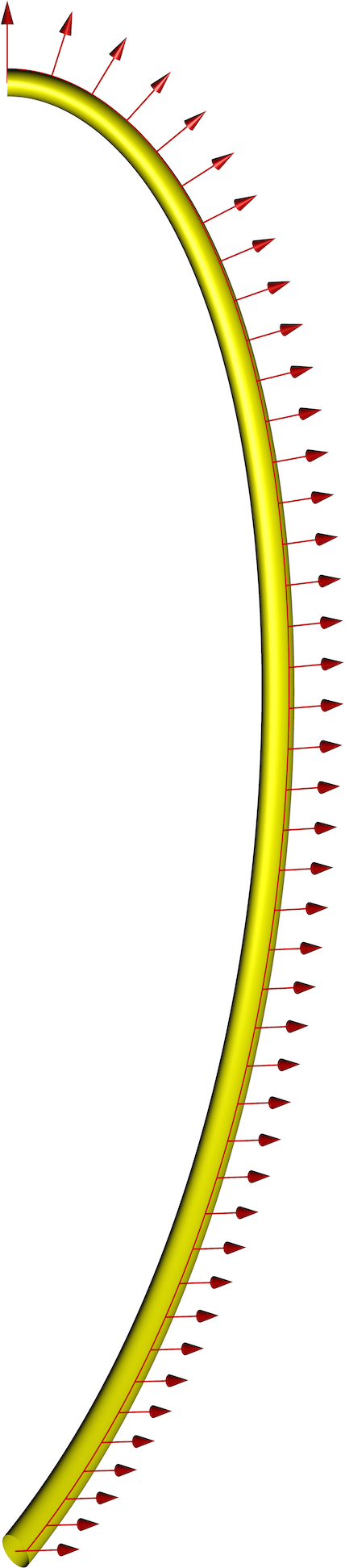}
\hspace{0.03\textwidth}(b)\hspace{0.01\textwidth}
\includegraphics[scale=1.8,valign=t]{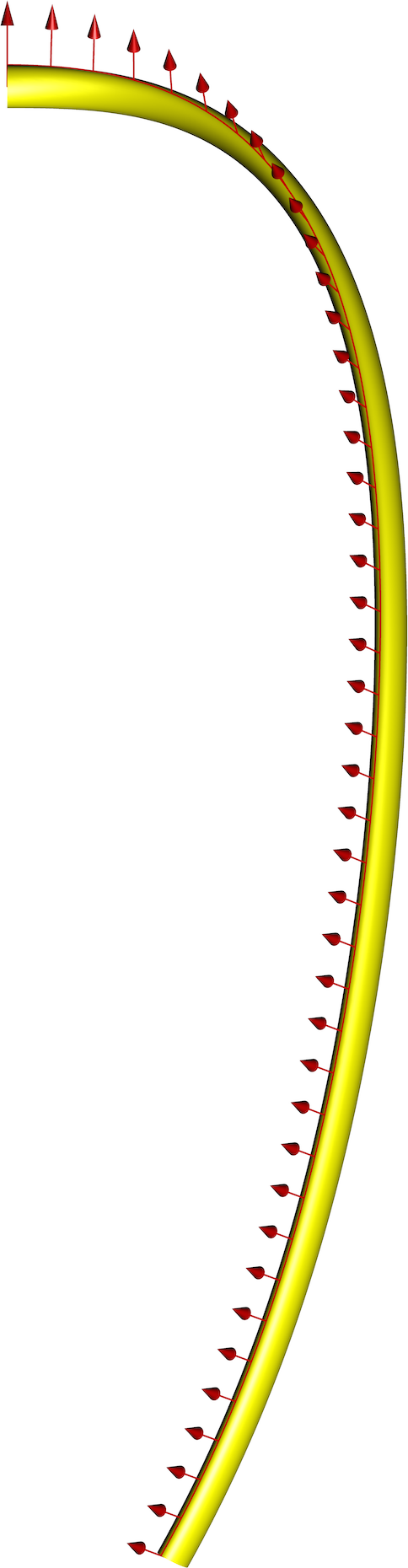}
\hspace{0.06\textwidth}
\includegraphics[width=0.46\textwidth,valign=t]{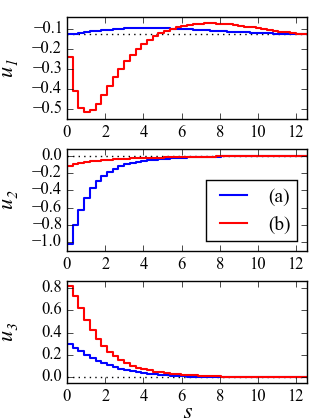}

\caption{Flexible beams with a curvilinear intrinsic shape that are clamped at one end relax, due to their own weight, towards configurations in which all of the strain fields depart from their intrinsic values, marked with dotted lines in the graph. We compare the two cases in which the intrinsic bending is in the direction of higher flexural rigidity (a) with $a_1/a_2=10$ and lower flexural rigidity (b) with $a_1/a_2=0.1$. In both cases the twisting rigidity $a_3$ equals the smaller of the two flexural rigidities, and shearing and stretching are strongly penalized. 
The deviation from the intrinsic shape is clearly stronger close to the clamped end and fades out towards the free end. Notably, the flexural density $u_1$ is everywhere hindered compared to its intrinsic value in case (a), whereas in case (b) it is accentuated in the region closer to the clamped end as a consequence of the different flexural rigidities and the greater amount of twisting.}\label{fig:beams}
\end{figure}

In our second example, we consider rods of relaxed length $L$ clamped only at one end and subject to their own weight. 
The effect of the weight is taken into account by adding to the energy $\E$ the term
\begin{equation}
\E_{\vc g}=-\int_\Omega \rho(s,\zeta_1,\zeta_2)\vc g\cdot\vc p(s,\zeta_1,\zeta_2)\,d(s,\zeta_1,\zeta_2),
\end{equation}
where the vector $\vc g$ is the gravitational acceleration, $\rho$ is the mass density of the rod, and $\vc p$ is defined according to \eqref{eq:p}.
In the discrete setting, the weight is uniformly distributed and acts effectively as point loads at the barycenters of the discrete cross sections.

The intrinsic shapes of these rods feature a uniform flexural density with $\bar{u}_1=-\pi/2L$ (they span a quarter of a circle) and the other strain fields have uniform values  $\bar{u}_2=\bar{u}_3=\bar{v}_1=\bar{v}_2=0$ and $\bar{v}_3=1$. We compare the two cases in which the intrinsic bending is in the direction of higher flexural rigidity with $a_1/a_2=10$ (Figure~\ref{fig:beams}a) and lower flexural rigidity with $a_1/a_2=0.1$ (Figure~\ref{fig:beams}b). In both cases the twisting rigidity $a_3$ equals the smaller of the two flexural rigidities. Here and in the next example shearing and stretching are strongly penalized, with $b_1/a_3=b_2/a_3=b_3/a_3=10^4$. The load generated by weight in combination with the curved intrinsic shapes produces eqilibria in which all of the strain fields must depart from their intrinsic values to effectively balance the load (Figure~\ref{fig:beams}, right).

\begin{figure}
\centering

(a)\hspace{0.12\textwidth}
\includegraphics[width=0.19\textwidth]{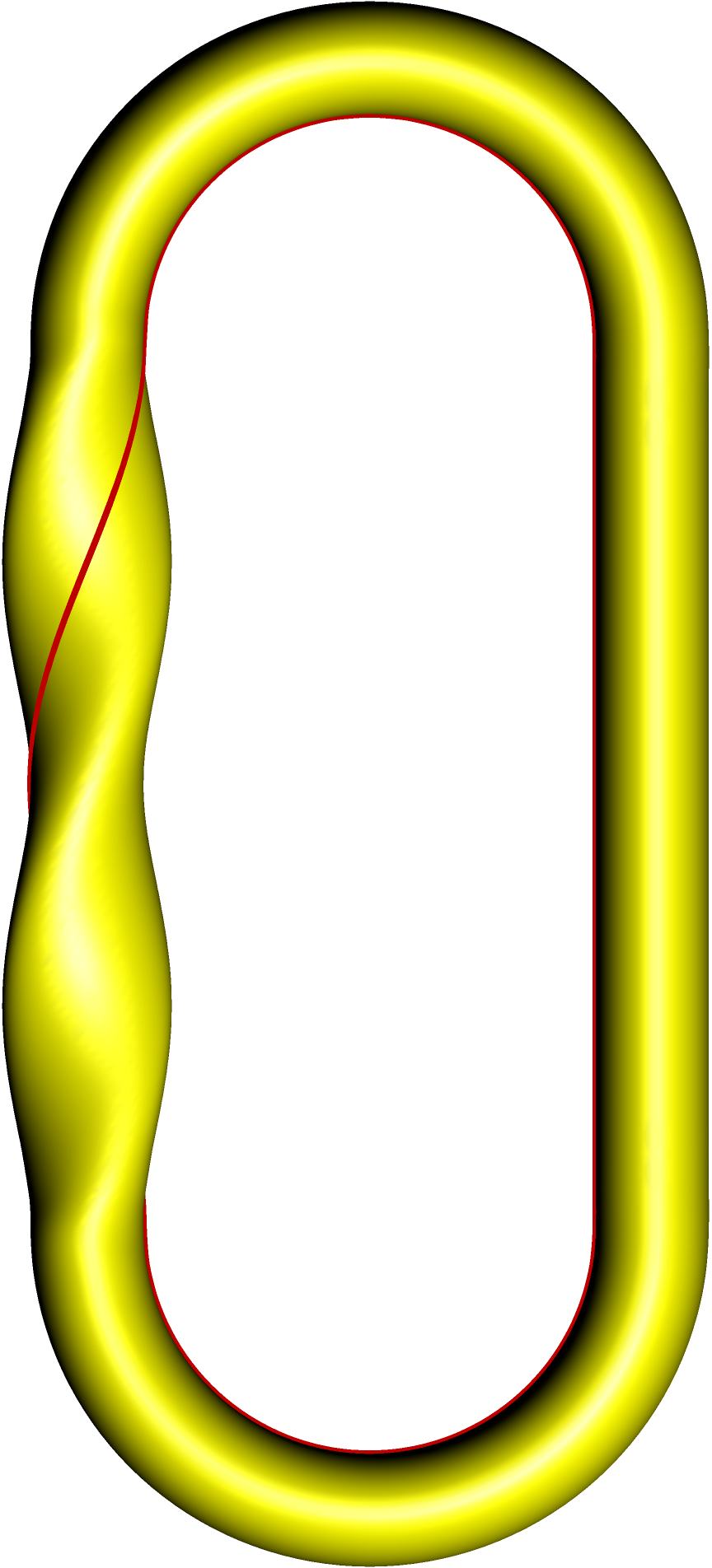}
\hspace{0.15\textwidth}(b)\hspace{0.0\textwidth}
\includegraphics[width=0.40\textwidth]{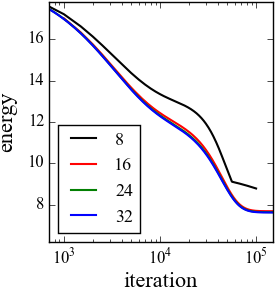}
\hspace{0.06\textwidth}

\vspace{8mm}

(c)
\includegraphics[width=0.46\textwidth]{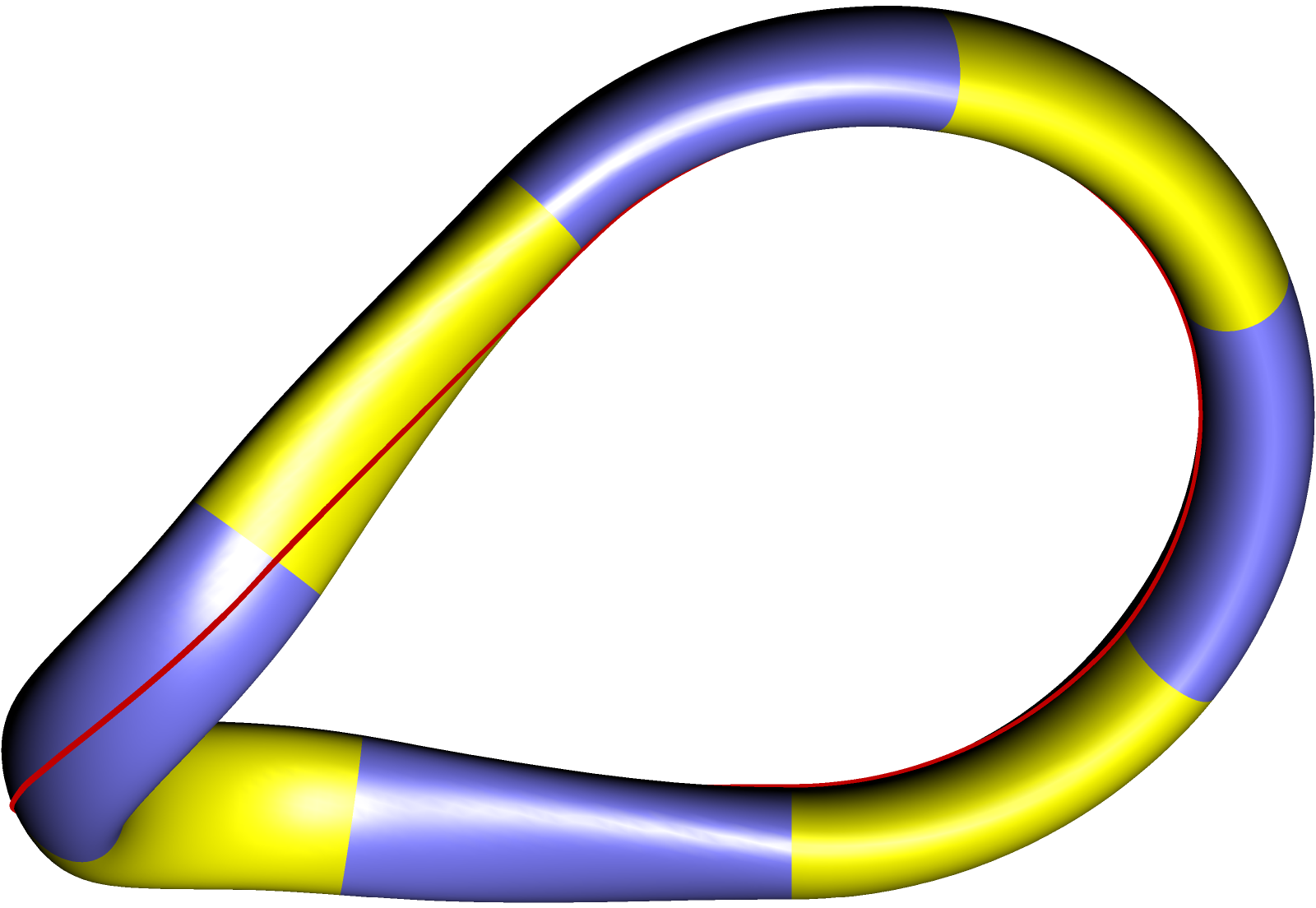}
(d)
\includegraphics[width=0.46\textwidth]{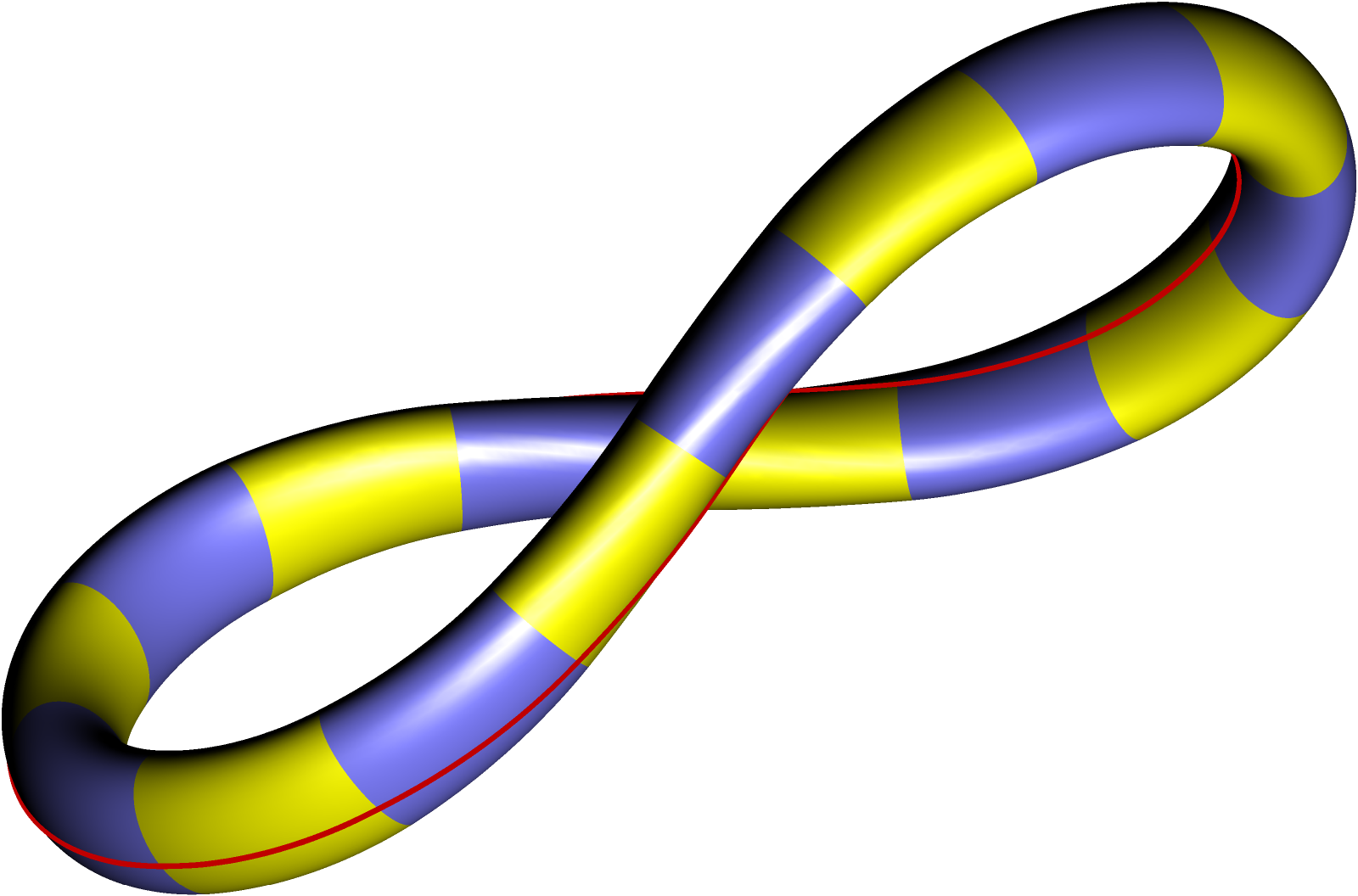}

\vspace{8mm}

(e)
\includegraphics[width=0.46\textwidth]{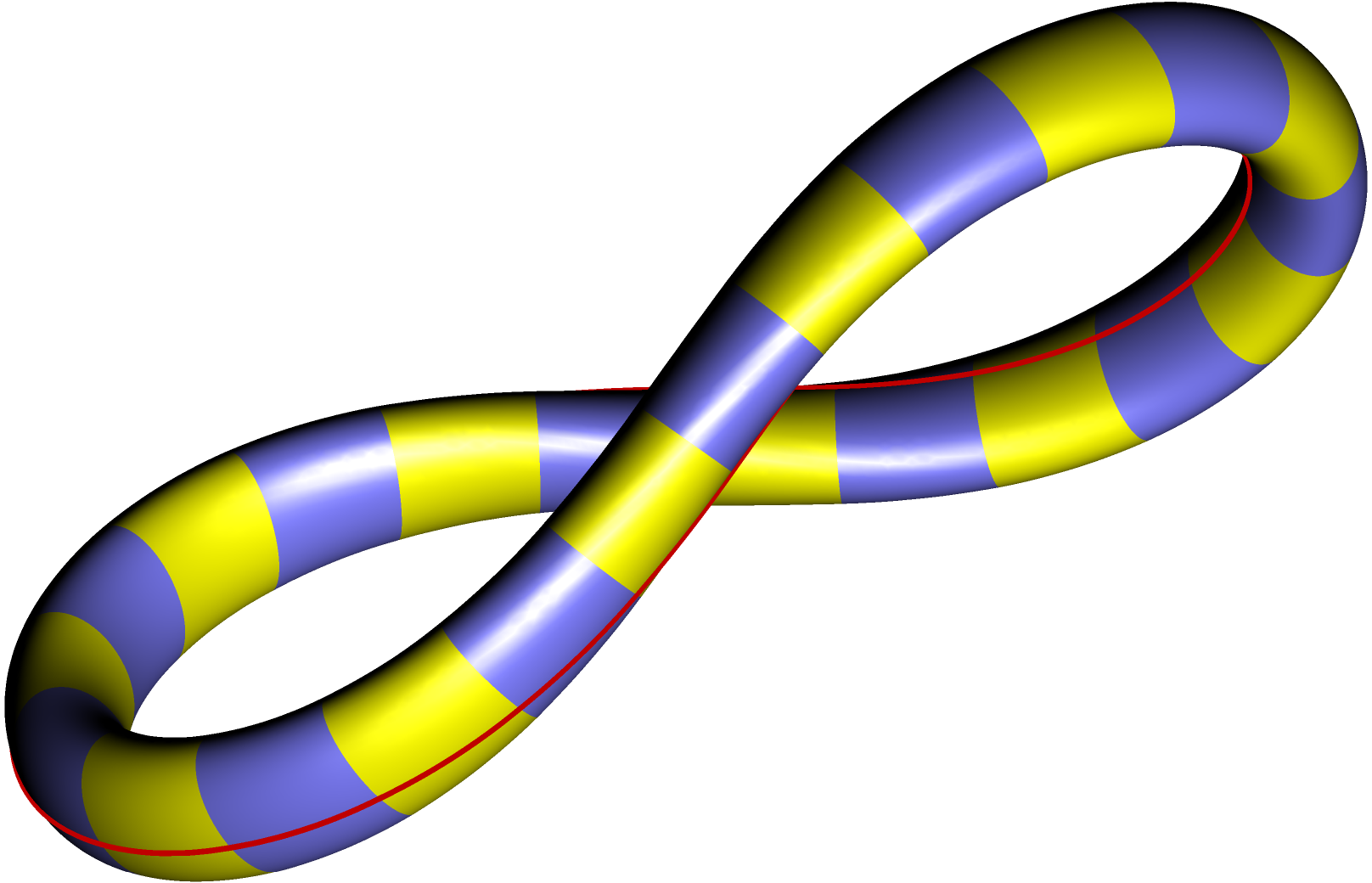}
(f)
\includegraphics[width=0.46\textwidth]{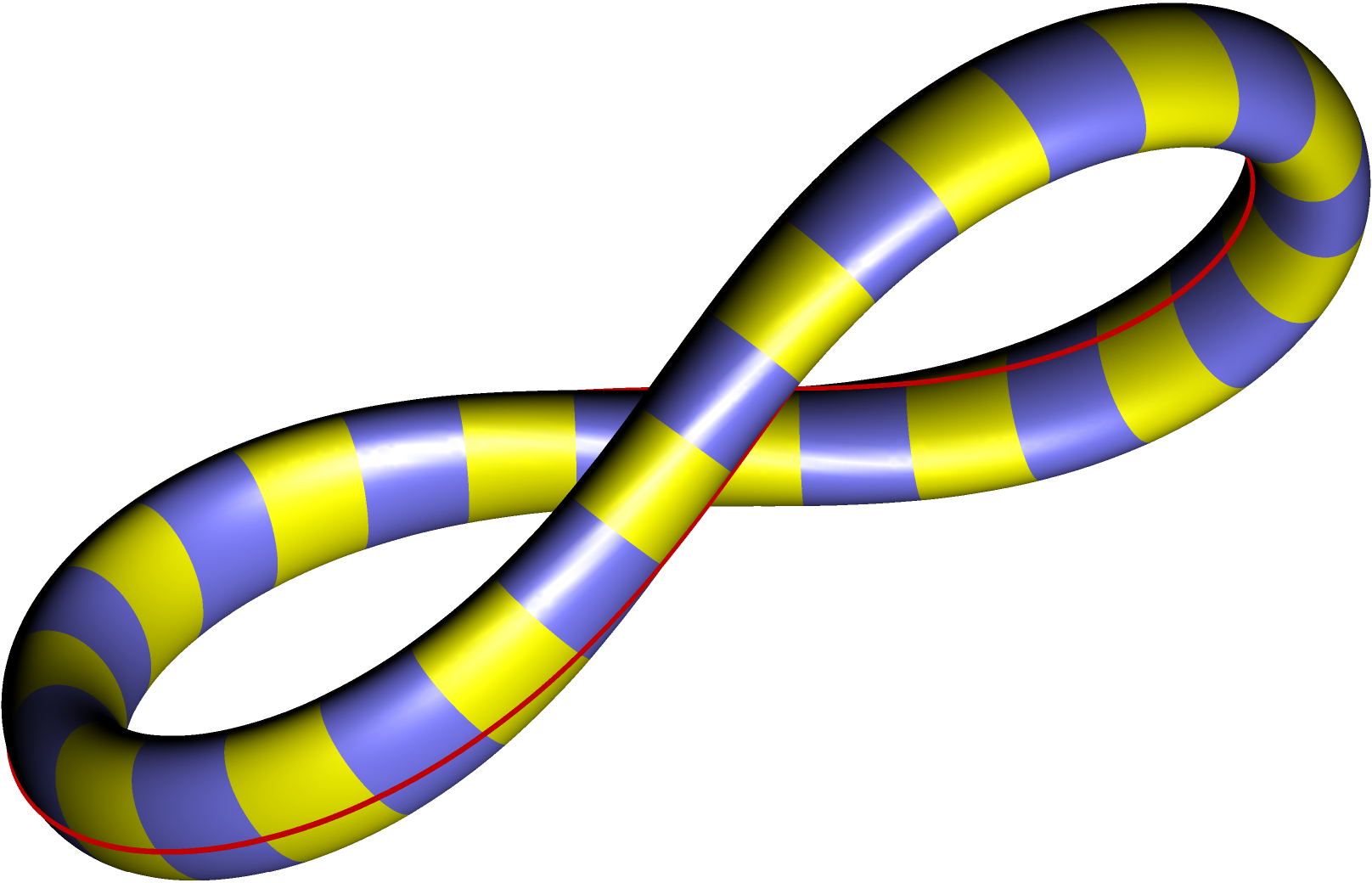}

\caption{Frustrated equilibrium shapes can be studied with a rather coarse discretization. 
A rod of length $L=2\pi$ with straight and twist-free intrinsic shape is forced by the clamping conditions into forming a closed loop with total twist $T=2\pi$. In the initial configuration (a) for the energy minimization algorithm flexural strains are confined to two opposite quarters of the loop and twisting is concentrated in a third sector. The rod is discretized by using 8, 16, 24, and 32 elements. The energy relaxation curves are depicted in panel (b). The approximation with 8 elements suffers from a locking phenomenon, that prevents a complete relaxation, and reaches a configuration (c) far from a realistic equilibrium. On the other hand, with just 16 elements it is possible to reach an equilibrium configuration (d) extremely close to those obtained with 24 (e) and 32 elements (f). The corresponding strain fields are shown in Figure~\ref{fig:loop-plot}.}\label{fig:loop}
\end{figure}

In the last example, we neglect the effects of weight, but we use clamping conditions at both ends to generate a highly frustrated equilibrium shape. We consider a rod of length $L=2\pi$ with straight and twist-free intrinsic shape. This configuration is unattainable since the clamping conditions force the rod into forming a closed loop with total twist $T=2\pi$, so the system relaxes towards a nontrivial equilibrium configuration. As initial condition for the gradient flow algorithm we assign a shape in which flexural strains are confined to two opposite quarters of the loop and twisting is concentrated in a third sector, as depicted in Figure~\ref{fig:loop}a. 
The elliptical cross section translates in anisotropic flexural rigidities given by $a_1/a_2=10$, while $a_3/a_2=1$. 

\begin{figure}
\centering

\includegraphics[width=\textwidth]{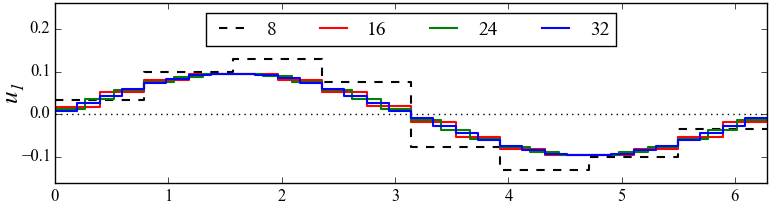}

\includegraphics[width=\textwidth]{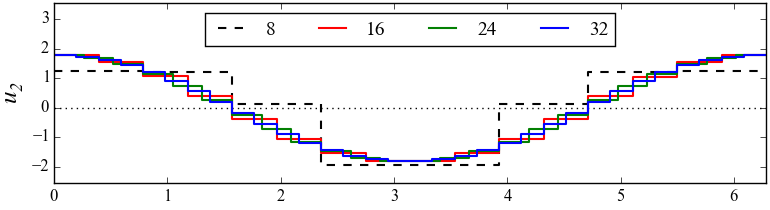}

\includegraphics[width=\textwidth]{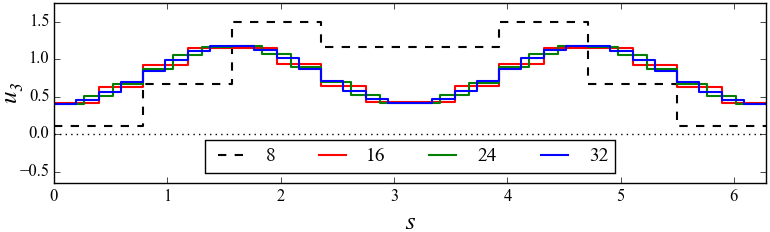}

\caption{Strain fields relative to the relaxed configurations depicted in Figure~\ref{fig:loop}. 
The case of 8 elements (dashed curves) remains rather far from the other cases, which nicely collapse providing increasingly better approximations of the same curves. In particular, the twisting strain $u_3$ remains significantly more localized in the case with only 8 elements, while it is distributed with two clearly separated peaks in the other cases. 
In all cases, the strains remain far from the vanishing intrinsic values, marked with a dotted line, showing the frustrated nature of these equilibria.}\label{fig:loop-plot}
\end{figure}

To evaluate the effect of the discretization, we approximate the rod by using 8, 16, 24, and 32 elements. The results of the relaxation algorithm are presented in Figure~\ref{fig:loop} and Figure~\ref{fig:loop-plot}. The crudest approximation with 8 elements suffers from a locking effect, preventing the complete relaxation of the system. In this case, we can observe a kink in the relaxation curve at about $5\times10^4$ iterations, which is absent in the relaxation curve for the cases with more elements (Figure~\ref{fig:loop}b). After that point, the rod is stuck in the configuration depicted in Figure~\ref{fig:loop}c, and the further spurious decrease in energy that is observed is due to the increasing error relative to the closure constraint. On the other hand, with just 16 elements we obtain a tangible energy relaxation towards an equilibrium configuration (Figure~\ref{fig:loop}d) extremely close to those obtained with 24 and 32 elements (Figure~\ref{fig:loop}e-f). By analyzing in more detail the strain fields of the relaxed configurations, we can indeed observe that the case of 8 elements remains rather far from the other cases, which nicely collapse providing increasingly better approximations of the same curves (Figure~\ref{fig:loop-plot}). In particular, the twisting strain $u_3$ remains significantly more localized in the case with only 8 elements, while it is distributed with two clearly separated peaks in the other cases. An important difference between the case of anisotropic flexural rigidity and the more classical case of equal flexural rigidities is the fact that the twisting strain in the equilibrium configuration is not uniformly distributed along the rod.

\section{Framed curves}\label{sec:framed-curve}

In this section, we derive a description of framed curves as degenerate rods. This provides an appealing way to discern the nature of the geometric invariants associated with a regular curve, namely a curve for which the tangent field is well-defined and everywhere continuous. Although Bishop~\cite{Bis75} carried out a careful analysis of this subject more than forty years ago, we show that his construction is valid under weaker assumptions, made clear by our alternate derivation. Moreover, we believe that our analysis provides additional compelling motivations for the exclusive use of parallel adapted frames for describing the kinematics of curves. 

We view framed curves as rods with cross sections that shrink to single points. Those cross-sections are thus clearly invariant under rotations. This degeneracy is reflected by the loss of meaning of the shear and twisting parameters. 
Indeed, to bring a cross section at $s$ (a point in space) onto an adjacent cross section at $s+ds$ it is necessary only to adjust the direction of the movement, as specified by assigning $u_1(s)$ and $u_2(s)$, and the intensity of the movement, as specified by $v_3(s)$. This suffices to rigidly move from one point to another, thus making unnecessary the use of $v_1(s)$, $v_2(s)$, or $u_3(s)$.

In this degenerate setting, it seems reasonable to identify the curve traced by the degenerate cross sections with the base curve, which clearly acquires a prominent role. It then becomes possible to exploit the freedom accorded to us by the degeneracy to definite a frame that is---as much as possible---generated by the geometry of the base curve.
Since shearing and twisting are no longer meaningful, we set, for every $s\in[0,L]$, $v_1(s)=v_2(s)=u_3(s)=0$  in \eqref{eq:Cauchy} and we identify $\vc d_3$ with the normalized tangent field $\vc t$ to the base curve, obtaining
\begin{equation}\label{eq:Cauchy-FC}
\left\{\begin{aligned}
&\vc x'(s)=v_3(s)\vc t(s),\\
&\vc t'(s)=u_2(s)\vc d_1(s)-u_1(s)\vc d_2(s),\\
&\vc d'_1(s)=-u_2(s)\vc t(s),\\
&\vc d'_2(s)=u_1(s)\vc t(s).
\end{aligned}\right.
\end{equation}
Extension and contraction, being simply based on distances between points, remain meaningful. For this reason, $v_3$ is not set identically equal to unity and, thus, the parameter $s$ is not necessarily the arclength along the base curve. Nevertheless, the physical requirement ruling out the interpenetration of matter forces the condition $v_3>0$. 
The solution of \eqref{eq:Cauchy-FC} would provide the curve (which is regular if we assume that $v_3$ is a positive and continuous field) and a \emph{relatively parallel adapted frame} (in the language of Bishop~\cite{Bis75}), sometimes referred to as natural frame, inertial frame, rotation-minimizing frame, or Fermi--Walker frame.

It is important to clarify to what extent the geometry of a curve determines such a frame. To this end, we seek to express the fields $v_3$, $u_1$, and $u_2$ in terms of the field $\vc x$ and its derivatives. We immediately obtain $v_3=|\vc x'|$ \red{and the standard expressions}
\begin{equation}
\vc t=\frac{\vc x'}{v_3}\qquad\text{and}\qquad
\vc t'=\frac{1}{v_3}\bigg(\vc x''-\frac{\vc x''\cdot\vc x'}{v_3^2}\vc x'\bigg)
\end{equation}
\red{for $\vc t$ and $\vc t'$ in terms of $\vc x'$ and $\vc x''$.}
If we now consider the integral forms of \eqref{eq:Cauchy-FC}$_{3,4}$, namely
\begin{equation}\label{eq:int-form}
\left\{\begin{aligned}
\vc d_1(s)=\vc d_1(0)-\int_0^s u_2(r)\vc t(r)\,dr,\\
\vc d_2(s)=\vc d_2(0)+\int_0^s u_1(r)\vc t(r)\,dr,
\end{aligned}\right.
\end{equation}
together with \eqref{eq:Cauchy-FC}$_{2}$, 
we can take the scalar product of \eqref{eq:int-form}$_{1,2}$ with $\vc t'$ to give
\begin{equation}\label{eq:Volterra}
\left\{\begin{aligned}
u_2(s)=\vc d_1(0)\cdot\vc t'(s)-\int_0^s u_2(r)\vc t(r)\cdot\vc t'(s)\,dr,\\
u_1(s)=\vc d_2(0)\cdot\vc t'(s)+\int_0^s u_1(r)\vc t(r)\cdot\vc t'(s)\,dr.
\end{aligned}\right.
\end{equation}
The integral equations \eqref{eq:Volterra} are Volterra equations of the second kind and, as such, admit a unique solution which is continuous on the interval $[0,L]$. (See, for example, Kress~\cite[Theorem 3.10]{Kress2014}.)

It is now worth commenting on the regularity needed for the forgoing construction. In particular, the fields $u_1$ and $u_2$ need not be continuous for \eqref{eq:Cauchy-FC} to have a unique solution. Correspondingly, granted that $\vc t'$ is a square-integrable field, the solutions $u_2$ and $u_1$ of \eqref{eq:Volterra} are also square-integrable fields, as discussed by Tricomi~\cite{Tricomi1957}. A choice that is convenient for most practical purposes is to view $u_1$, $u_2$, and $\vc t'$ as piecewise-continuous fields, but weaker regularity is also allowed.
We have thus shown that the prescription of a continuously-differentiable curve $\vc x:[0,L]\to\R^3$, such that $|\vc x'(s)|>0$ for any $s$ in $[0,L]$, together with a choice for the value $\vc{d}_1(0)$ of the material director $\vc d_1$ at one end of the curve (since $\vc d_2(0)$ is simply given by $\vc t(0)\times\vc d_1(0)$) uniquely determines the scalar fields $v_3$, $u_1$, and $u_2$ that, in turn, suffice to build a relatively parallel adapted frame by solving \eqref{eq:Cauchy-FC}. 

Since cross sections are here reduced to a single point, the director $\vc d_1(0)$ can be arbitrarily chosen in the plane normal to $\vc x'(0)$. 
As noted by Bishop~\cite{Bis75}, due to this arbitrariness, for any regular curve there exists a one-parameter family of relatively parallel adapted frames that are completely determined by the geometry of the curve. The fields $u_1/v_3$ and $u_2/v_3$ are a parametrization of the \emph{normal development} of the curve, but they are not invariant under rotations of $\vc d_1(0)$ in the normal plane at $s=0$.
The shape of the graph of the normal development in the product of its centro-Euclidean plane with the interval $[0,L]$ is the true geometric invariant of the curve. If we introduce the scalar fields $\kappa$ and $\theta$ through the identification
\begin{equation}\label{eq:complexify}
\kappa(s)e^{\mathrm{i}\theta(s)}:=\frac{u_2(s)+\mathrm{i}u_1(s)}{v_3(s)}
\end{equation}
(introduced by Hasimoto~\cite{Has72} to describe vortex filaments), then the two scalar fields representing the geometric invariants of the curve are the \emph{square-integrable} curvature field $\kappa$ and the \emph{measure-valued} torsion field $\tau:=\theta'$.
It is important to observe that, even though the phase $\theta$ itself is not defined where $\kappa$ vanishes, its derivative  $\tau$ remains well-defined in the sense of distributions, granted that we simply set $\theta(s)=0$ when $\kappa(s)=0$.

It is possible to state in terms of the strain fields $u_1$ and $u_2$ (assuming $v_3$ equal to unity) also the famous problem raised independently by Efimov~\cite{Efi47} and Fenchel~\cite{Fen51} of identifying necessary and sufficient conditions on curvature and torsion for the curve parametrized by $\vc x$ to be closed. This amounts to requiring that a suitable restriction of the operator $\uop(L;0)$ defined in \eqref{eq:uop} be equal to the identity map. If we simply want the curve to be closed, we consider only the action of $\uop(L;0)$ on the components of the field $\vc x$. If we require a smooth closure, we then consider the restriction of $\uop(L;0)$ to the components of $\vc x$ and $\vc t=\vc d_3$. This general solution of the closed curve problem is identical in spirit to that provided by Schmeidler~\cite{Sch56} (and later by Hwang~\cite{Hwa81}) in terms of continuous curvature and torsion fields, but it readily shows that the results extend to the case of square-integrable curvature and measure-valued torsion. As we previously observed, an explicit expression of $\uop(L;0)$ in terms of the strain fields is available only in very special cases, limiting the scope of applicability of the general closure conditions.

Consistent with our construction of framed curves and with the emphasis on the curve geometry appropriate to these objects, we emphasize the importance of maintaining a clear distinction between framed curves and special Cosserat rods.
\red{Even though the two concepts have been successfully combined in the context of Kirchhoff rods} (see, for instance, the papers of Langer \& Singer~\cite{LanSin96} and Goriely \& Tabor~\cite{GorTab00}, the models for DNA reviewed by Swigon~\cite{Swi09}, and the recent contributions by Kawakubo~\cite{Kaw08}), \red{this was possible because relatively parallel adapted frames were correctly used to describe the geometry of the base curve, while an additional material frame was employed to keep track of the mechanical twist.
Nevertheless, we have shown that the geometry of the base curve is not a necessary starting point to define the shape of a rod, since the strain fields $u_i$ and $v_i$, $i=1,2,3$, are the only degrees of freedom needed to characterize that shape.}
In summary, it seems more appropriate to explicitly use rod theory when dealing with mechanical models (as exemplified by the works of Domokos~\cite{Dom95} and Domokos \& Healey~\cite{DomHea01}, and many others cited above) and framed curves when focusing on geometry (as exemplified by the works of Starostin \& van der Heijden~\cite{StavdH15}, Bohr \& Markvorsen~\cite{BohMar16}, da Silva~\cite{daS16}, and Honda \& Takahashi \cite{HonTak16}).

\red{We finally stress that, even in treatments of purely geometrical questions connected to space curves, the classical Frenet frame is not a suitable tool for two important reasons. First, whenever a curve has a straight portion, the curvature $\kappa$ vanishes and the Frenet normal is not defined, even though the geometric invariants $\kappa$ and $\tau$ are well-defined everywhere. Second, even when $\kappa>0$ at all points of a curve, it may be no more than square-integrable (with a measure-valued $\tau$), so that the corresponding Frenet frame could possibly be discontinuous.
(Consider, for example, the properties of the base curve of a M\"obius band, as described by Randrup \& R{\o}gen~\cite{RanRog96}.)
We thus see that the family of relatively parallel adapted frames, being uniquely determined by the geometric invariants of a regular curve and containing only globally-defined and continuous frames, should always be preferred over the Frenet frame. This, however, should not obscure the fact that $\kappa$ and $\tau$ are the true geometric invariants of the curve, while $u_1$, $u_2$, and $v_3$ provide a convenient parametrization of the shape of the curve, having selected $\vc d_1(0)$.}

\section{Conclusions}\label{sec:conclusions}

We have described how the essential degrees of freedom that encode the shape of a special Cosserat rod, namely those features that are invariant under isometries of the three-dimensional ambient space, correspond to a path traced in the special Euclidean algebra. 
The typical regularity of such path, relevant for physical applications, is that of a square-integrable map.
This representation of the shape is intrinsic to the description of the mechanical properties of a rod, since it is given directly in terms of the strain fields that underpin the elastic response of special Cosserat rods.

The Lie algebraic description of the rod shapes leads to an appealing discretization scheme that can be successfully applied to the analysis of shape relaxation problems under strongly nonlinear and non-convex geometric constraints, such as the clamped-ends and closure requirements.

We have recovered the notion of a framed curve as a Cosserat rod with point-like cross sections. 
That degeneracy is reflected on the actual degrees of freedom of the system. 
From this standpoint, we have highlighted the essential difference between rods and framed curves, and we have clarified why the family of relatively parallel adapted frames is not suitable for describing the mechanics of rods but it is the appropriate tool for dealing with the geometry of curves.

\vspace{4mm}
\noindent
\textbf{Funding:} The authors acknowledge support from the Okinawa Institute of Science and Technology Graduate University with subsidy funding from the Cabinet Office, Government of Japan.



\end{document}